\documentclass[aps,prd,nofootinbib,groupedaddress,twocolumn,floatfix]{revtex4}
\usepackage{graphicx}
\usepackage{epsfig}
\usepackage{dcolumn}
\usepackage{amsmath}
\def\Journal#1#2#3#4{{#1} {\bf#2}, #3 (#4)}

\def\NPB{{\rm Nucl. Phys.} B}
\def\PLB{{\rm Phys. Lett.}  B}
\def\PRL{\rm Phys. Rev. Lett.}
\def\PRD{{\rm Phys. Rev.} D}

\def\ep{\epsilon}

\def\la{\langle}
\def\ra{\rangle}

\def\be{\begin{equation}}
\def\ee{\end{equation}}
\def\bea{\begin{eqnarray}}
\def\eea{\end{eqnarray}}
\input{epsfig.sty}
\begin{document}
\title{Distribution amplitudes and decay constants for 
$(\pi,K,\rho,K^*)$ mesons in light-front quark model}
\author{ Ho-Meoyng Choi$^{a}$ and Chueng-Ryong Ji$^{b}$\\
$^a$ Department of Physics, Teachers College, Kyungpook National University,
     Daegu, Korea 702-701\\
$^b$ Department of Physics, North Carolina State University,
Raleigh, NC 27695-8202}
\begin{abstract}
We present a calculation of the quark distribution amplitudes(DAs), 
the Gegenbauer moments, and decay constants for $\pi,\rho,K$ and $K^*$ 
mesons using the light-front quark model. While the quark DA for $\pi$ 
is somewhat broader than the asymptotic one, that for $\rho$ meson
is very close to the asymptotic one. The quark DAs for $K$ and $K^*$
show asymmetric form due to the flavor SU(3)-symmetry breaking effect.
The decay constants for the transversely polarized $\rho$ and $K^*$
mesons($f^T_\rho$ and $f^T_{K^*}$) as well as the longitudinally
polarized ones($f_\rho$ and $f_{K^*}$) are also obtained. 
Our averaged values for $f^T_V/f_V$, i.e. 
$(f^T_\rho/f_\rho)_{\rm av}=0.78$ and $(f^T_{K^*}/f_{K^*})_{\rm av}=0.84$,
are found to be consistent with other model predictions.
Especially, our results for the decay constants are in a good agreement with 
the SU(6) symmetry relation, $f^T_{\rho(K^*)}=(f_{\pi(K)}+f_{\rho(K^*)})/2$.
\end{abstract}


\maketitle
\section{Introduction}

Hadronic distribution amplitudes(DAs)
are important ingredients in applying QCD to hard 
exclusive processes via the factorization 
theorem~\cite{BL,ER,CZ}. They provide essential
information on the nonperturbative structure of hadron
describing the distribution of partons 
in terms of the longitudinal momentum fractions 
inside the hadron. Both the 
electromagnetic form factors at high
$Q^2$ and the $B$-physics phenomenology are highly relevant
to the detailed computation of hadronic DAs. 


During the past few decades, there have been many theoretical efforts
to calculate the pion DA using nonperturbative
methods such as the QCD sum rule~\cite{CZ,BZ,Agaev,SY,BMS1,BMS2},
lattice calculation~\cite{Ali,DS,QCDSF,Go,Del}, chiral quark model from
the instanton vacuum~\cite{PPRWG,NKHM,ADT}, Nambu-Jona-Lasinio(NJL)
model~\cite{Ar,PR}, and light-front quark model(LFQM)~\cite{JC1,JC2}.
It is well known that the shape of the pion quark DA is very important 
in the predictions of pion electromagnetic form factor in both 
nonperturbative and perturbative momentum transfer regimes. 
The QCD sum-rule based analysis~\cite{SY,BMS1} of
the  $\pi-\gamma$ transition form factor $F^{\pi\gamma\gamma^*}(Q^2)$
measured by  the CLEO experiment~\cite{Gron}
has shown that neither double-humped DA for the pion predicted by Chernyak and
Zhitnitsky~\cite{CZ} nor the
asymptotic one are favored at the 2$\sigma$ level of accuracy.
It is also interesting to note that the recent anti-de Sitter space 
geometry/conformal field theory(AdS/CFT)
prediction~\cite{Ads} for the meson DA is 
$\phi_{\rm AdS/CFT}(x)\propto\sqrt{x(1-x)}$, which would approach to 
the asymptotic form $x(1-x)$ only in the limit of $\ln Q^2\to\infty$. 
The shape of $\phi_{\rm AdS/CFT}(x)$ increases the usual perturbative 
QCD (PQCD) predictions for the pion form factor and $\pi-\gamma$ 
transition form factor
by 16/9 and 4/3, respectively. 
In our recent LFQM application to the PQCD analysis of 
the pion form factor~\cite{Conformal}, we further found a correlation 
between the shape of quark DA and the amount of soft and hard contributions 
to the pion form factor. Similar to the previous findings from the Sudakov 
suppression of the soft contribution 
(or enhancement of the hard contribution)~\cite{LS,Jacob,JK,Doro},
our results indicated that
the suppression of the endpoint region for the quark DA corresponds to the
suppression(enhancement) of the soft(hard) contribution.

Another important area that requires detailed study of meson DAs is 
the B-physics phenomenology under intense experimental investigation at 
BaBar
and Belle experiments. The $K,\rho$ and $K^*$ DAs have attracted attention
rather recently~\cite{KMM,BBL,BZ2,Yang} due to the deep relevance to the 
exclusive $B$-meson decays to $(K,\rho,K^*)$ mesons. In particular, 
the $SU(3)$ flavor symmetry breaking effect in the meson quark
DA including strange quark is important for the predictions of 
exclusive $B_{u,d,s}$-decays to light pseudoscalar and vector mesons
in the context of CP-violation and Cabibbo-Kobayashi-Maskawa quark
mixing matrix studies. The $SU(3)$ breaking effect is realized in the
difference between the longitunal momenta of the strange and nonstrange
quark, $\la x_s-x_{u(d)}\ra\neq 0$, in the two particle Fock components
of the meson. The similar effect was also found in our 
PQCD analysis~\cite{DD} for the exclusive heavy meson pair production in
$e^+e^-$ annihilations at $\sqrt{s}=10.6$ GeV. 
Not only the shape of the heavy meson quark DAs matters 
in the prediction of the cross section for the heavy meson pair productions,
but also the cross section ratios for 
$\sigma(e^+e^-\to D^+_s D^-_s)/\sigma(e^+e^-\to D^+D^-)$
and $\sigma(e^+e^-\to B^0_s {\bar B}^0_s)/\sigma(e^+e^-\to B^+B^-)$
deviate from 1 appreciably due to the $SU(3)$ symmetry breaking. 


A particularly convenient and intuitive framework in applying PQCD to 
exclusive processes is based upon the light-front(LF) Fock-state 
decomposition of hadronic state.
In the LF framework, the valence quark DA is computed from the valence
LF wave function
$\Psi_n(x_i,{\bf k}_{\perp i})$
of the hadron at equal LF time $\tau=t + z/c$ which
is the probability amplitude to find $n$
constituents(quarks,antiquarks, and gluons) with LF momenta
$k_i=(x_i,{\bf k}_{\perp i})$ in a hadron. Here, $x_i$ and ${\bf k}_{\perp i}$
are the LF momentum fraction and the transverse momenta of the $i$th
constituent in the $n$-particle Fock-state, respectively.
If the factorization theorem in PQCD is applicable to
exclusive processes, then the invariant amplitude ${\cal M}$ for 
exclusive process factorizes into the convolution of the process-independent 
valence quark DA $\phi(x,\mu)$ with the process-dependent hard scattering
amplitude $T_H$~\cite{BL}, i.e. 
\bea\label{fac}
{\cal M}=\int[dx_i]\int[dy_i]\phi(x_i,\mu)T_H(x_i,y_i,\mu)\phi(y_i,\mu),
\eea
where $[dx_i]=\delta(1-\Sigma^n_{k=1}x_k)\Pi^n_{k=1}dx_k$ and $n$ is the
number of quarks in the valence Fock state. 
Here, $\mu$ denotes the separation scale between perturbative and 
nonperturbative regime.
Since the collinear divergences
are summed in $\phi(x_i,\mu)$, the hard scattering amplitude $T_H$ can be
systematically computed as a perturbative expansion in $\alpha_s(\mu)$.
To implement the factorization theorem given by Eq.~(\ref{fac}) at high 
momentum transfer, the hadronic
wave function plays an important role linking between long distance
nonperturbative QCD encoded in DA and short distance PQCD encoded in $T_H$.

The quark DA of a meson, $\phi(x,\mu)$, is the probability of finding
collinear quarks up to the scale $\mu$ in the $L_z=0$($s$-wave) projection
of the meson wave function defined by
\bea\label{DA1}
\phi(x_i,\mu)=\int^{|{\bf k}_\perp|<\mu}[d^2{\bf k}_{\perp i}]
\Psi(x_i,{\bf k}_{\perp i}),
\eea
where 
\bea\label{kp}
[d^2{\bf k}_{\perp i}]=
2(2\pi)^3\delta\biggl[\sum^n_{j=1}{\bf k}_{\perp j}\biggr]
\Pi^n_{i=1}\frac{d^2{\bf k}_{\perp i}}{2(2\pi)^3}.
\eea 
Simple relativistic quark-model based on the LF framework
has been studied for various mesons\cite{Dziembowski,JC1,
JC2,DCJ}. Although the proof of duality between the LFQM and
the first principle QCD is not yet available,
we have attempted to fill the gap between the model wave function
and the QCD-motivated effective Hamiltonian\cite{Mixing,semi}.
The essential feature of our LFQM~\cite{Mixing,semi}
is to treat the gaussian radial wave
function as a trial function for the variational principle to the
QCD-motivated Hamiltonian. We saturate the Fock state expansion by the
constituent quark and antiquark, i.e. $H_{q\bar{q}}=H_0 + V_{\rm int}$,
where the interaction potential $V_{\rm int}$ consists of confining and
hyperfine interaction terms. From the variational principle minimizing
the central Hamiltonian with respect to the gaussian parameter,
we can find the optimum values of our model parameters and predict
the mass spectra for the low-lying ground state pseudoscalar and vector
mesons~\cite{Mixing,semi}.
We applied our LFQM for various exclusive processes such as
the electromagnetic form factors of $\pi$, $K$ and
$\rho$~\cite{Mixing,CJ_rho} mesons and semileptonic
and rare $B$ decays to $\pi$ and $K$~\cite{semi,CJ_K}. Our results for the
above exclusive processes were in a good agreement with the available data
as well as other theoretical model predictions. 

The purpose of this work is to calculate the quark DAs, the Gegenbauer
moments, and decay constants  
for $\pi,\rho,K$ and $K^*$ mesons using our LFQM and compare with other 
theoretical model predictions. 
As expected, while the odd Gegenbauer 
moments for $\pi$ and $\rho$ meson DAs are found to be zero due to 
isospin symmetry, the odd moments for $K$ and $K^*$ meson DAs are nonzero
due to the flavor SU(3)-symmetry breaking effect.
We compute the decay constants for the transversely polarized 
$\rho$ and $K^*$ mesons($f^T_\rho$ and $f^T_{K^*}$) as well as the 
longitudinally polarized ones($f_\rho$ and $f_{K^*}$) and compare with
the light-cone sum rule(LCSR) calculations, in which the ratio of 
$f^T_V$ and $f_V$ is an important ingredient for the LCSR predictions of
the $B\to\rho$ and $B\to K^*$ transition form factors. 
We also confirm that our results for the decay constants follow an old SU(6) 
symmetry relation~\cite{Leut}, $f^T_{\rho(K^*)}=(f_{\pi(K)}+f_{\rho(K^*)})/2$. 


The paper is organized as follows: In Sec.II, we briefly describe the 
formulation of our LFQM~\cite{Mixing,semi} and the procedure of fixing
the model parameters using the variational principle for the QCD-motivated 
effective Hamiltonian. The shape of the quark DA is then uniquely 
determined in our model calculation.  In Sec.III, the formulae for the 
quark DAs and decay constants of pseudoscalar and vector mesons are 
given in our LFQM. The Gegenbauer and $\xi(=x_1-x_2)$ moments are also 
given in this section. In Sec. IV, we present the numerical results for 
the decay constants, the quark DAs, the Gegenbauer and $\xi$ moments for 
($\pi, K,\rho, K^*$) mesons and compare with other theoretical model
predictions. Summary and conclusions follow in Sec.V. The relations
between $\xi$ and Gegenbauer moments are presented in Appendix A. 

\section{Model Description}
In our LFQM~\cite{Mixing,semi}, the meson wave function is given by
\bea\label{wf}
\Psi_M^{JJ_z}(x,{\bf k}_\perp,\lambda\bar{\lambda})
&=& \phi_R(x,{\bf k}_\perp)
{\cal R}^{JJ_z}_{\lambda\bar{\lambda}}(x,{\bf k}_\perp),
\eea
where $\phi_R(x,{\bf k}_\perp)$ is the radial wave function and
${\cal R}^{JJ_z}_{\lambda\bar{\lambda}}(x,{\bf k}_\perp)$ is the spin-orbit
wave function obtained by the interaction-independent Melosh
transformation~\cite{Melosh}
from the ordinary equal-time static spin-orbit wave function assigned by
the quantum numbers $J^{PC}$. The meson wave function in Eq.~(\ref{wf}) is
represented by the Lorentz-invariant variables $x_i=p^+_i/P^+$,
${\bf k}_{\perp i}={\bf p}_{\perp i}- x_i{\bf P}_\perp$ and $\lambda_i$,
where $P,p_i$ and $\lambda_i$ are the meson momentum, the momenta and the
helicities of the constituent quarks, respectively.

The radial wave function $\phi_R(x,{\bf k}_\perp)$ of a ground state
pseudoscalar meson($J^{PC}=0^{-+}$) is given by
\bea\label{phi}
\phi_R(x,{\bf k}_\perp)&=&\biggl(\frac{1}{\pi^{3/2}\beta^3}\biggr)^{1/2}
\exp(-{\vec k}^2/2\beta^2),
\eea
where ${\vec k}^2= {\bf k}^2_\perp + k^2_z$ and 
the gaussian parameter $\beta$ is related with the size of the meson.
Here, the longitudinal component $k_z$ of the three momentum is given
by $k_z=(x_1-\frac{1}{2})M_0 + (m^2_2-m^2_1)/2M_0$ with the invariant mass
\bea\label{M0}
M^2_0&=& \frac{{\bf k}^2_\perp + m^2_1}{x_1}
+\frac{{\bf k}^2_\perp + m^2_2}{x_2},
\eea
where $x_1=x$ and $x_2=1-x$.
The covariant form of the  spin-orbit wave functions for
pseudoscalar($J^{PC}=0^{-+}$) and vector($1^{--}$) mesons are given by
\bea\label{spin_cov}
{\cal R}^{00}_{\lambda\bar{\lambda}}&=&
-\frac{\bar{u}(p_1,\lambda)\gamma_5 v(p_2,\bar{\lambda})}
{\sqrt{2}[M^2_0-(m_1-m_2)^2]^{1/2}},
\nonumber\\
{\cal R}^{1J_3}_{\lambda\bar{\lambda}}&=&
-\frac{\bar{u}(p_1,\lambda)
\biggl[\not\!\ep(J_z) - \frac{\ep\cdot(p_1-p_2)}{M_0 + m_1 + m_2}\biggr]
v(p_2,\bar{\lambda})}
{\sqrt{2}[M^2_0-(m_1-m_2)^2]^{1/2}}.
\eea
The polarization vectors $\epsilon^\mu=(\ep^+,\ep^-,\ep_\perp)$
used in this analysis are given by
\bea{\label{pol_vec}}
\epsilon^\mu(\pm 1)&=&
\biggl[0,\frac{2}{P^+}{\bf\epsilon}_\perp(\pm)\cdot{\bf P_{\perp}},
{\bf\epsilon}_\perp(\pm 1)\biggr],
\nonumber\\
{\bf\epsilon}_\perp(\pm 1)&=&\mp\frac{(1,\pm i)}{\sqrt{2}},
\nonumber\\
\epsilon^\mu(0)&=&
\frac{1}{M_0}\biggl[P^+,\frac{{\bf P}^2_{\perp}-M^2_0}{P^+},
{\bf P}_{\perp}\biggr].
\eea
Note that $\sum_{\lambda\bar{\lambda}}R^{JJ_z\dagger}_{\lambda\bar{\lambda}}
R^{JJ_z}_{\lambda\bar{\lambda}}=1$.
The normalization of our wave function is given by
\bea\label{norm}
\sum_{\lambda\bar{\lambda}}&&\hspace{-0.5cm}
\int d^3 k|\Psi^{JJ_z}_M(x,{\bf k}_\perp,\lambda\bar{\lambda})|^2
\nonumber\\
&=&\int^1_0 dx
\int d^2{\bf k}_\perp\biggl(\frac{\partial k_z}{\partial x}\biggr)
|\phi_R(x,{\bf k}_\perp)|^2 = 1,
\eea
where the Jacobian of the variable transformation $\{x,{\bf k}_\perp\}\to
\vec{k}=({\bf k}_\perp,k_z)$ is given by
\bea\label{Jacob}
\frac{\partial k_z}{\partial x}=
\frac{M_{0}}{4x_1x_2}\biggl\{
1-\biggl[\frac{(m_1-m_2)^2}{M^2_{0}}\biggr]^2\biggr\}.
\eea
The effect of the Jacobi factor has been analyzed in Ref.~\cite{CJ_Jacob}.

The key idea in our LFQM~\cite{Mixing,semi} for mesons is to treat the radial
wave function $\phi_R(x,{\bf k}_\perp)$ as a trial function for the
variational principle to the QCD-motivated Hamiltonian saturating the
Fock state expansion by the constituent quark and antiquark.
The QCD-motivated effective Hamiltonian for a description of the meson
mass spectra is given by~\cite{GI}
\bea\label{Ham}
H_{q\bar{q}}=H_0 + V_{q\bar{q}}=
\sqrt{m^2_q + {\vec k}^2}
+ \sqrt{m^2_{\bar{q}} + {\vec k}^2} + V_{q\bar{q}}.
\eea
In our LFQM~\cite{Mixing,semi}, we use the two interaction potential 
$V_{q\bar{q}}$ for the pseudoscalar and vector mesons: (1) Coulomb
plus harmonic oscillator(HO), and (2) Coulomb plus linear confining 
potentials. In addition, the hyperfine interaction, which is essential to
distinguish vector from pseudoscalar mesons, is included for both cases,
viz.,
\bea\label{Vq}
V_{q\bar{q}}=V_0 + V_{\rm hyp}= a + {\cal V}_{\rm conf} - \frac{4\kappa}{3r}
+ \frac{2{\bf S}_q\cdot{\bf S}_{\bar{q}}}{3m_qm_{\bar{q}}}
\nabla^2V_{\rm Coul.},
\nonumber\\
\eea
where ${\cal V}_{\rm conf}=br(br^2)$ for the linear(HO) potential and
$\la{\bf S}_q\cdot{\bf S}_{\bar{q}}\ra=1/4(-3/4)$ for the vector(pseudoscalar)
meson.

We then take $\phi_R(x,{\bf k}_\perp)$ as our trial function to minimize the 
central Hamiltonian via
\bea\label{var}
\frac{\partial\la\Psi|[H_0 + V_0]|\Psi\ra}{\partial\beta}=0.
\eea
From the above constraint, only 4 parameters
are independent among the light-quark masses and the potential 
parameters $(m_q, \beta_{q\bar{q}},a,b,\kappa)(q=u,d)$. 
In order to determine these four parameters from the two experimental values of
$\rho$ and $\pi$ masses, we take the string tension $b=0.18$ GeV$^2$ and the constituent
$u$ and $d$ quark masses $m_u=m_d=0.22 (0.25) $ GeV for the linear (HO) potential, 
which are rather well known from
other quark model analyses commensurate with Regge phenomenology~\cite{GI}.
More detailed procedure of determining the model parameters of light quark
sector($u(d)$ and $s$) can be found in~\cite{Mixing,semi}. Our model parameters
for the light quark sector obtained by the variational principle are 
summarized in Table~\ref{t1}.
\begin{table}[t]
\caption{The constituent quark masses $m_{q}$(in GeV) and the gaussian
parameters $\beta_{q\bar{q}}$(in GeV) for the linear[HO]
potential obtained from the variational principle.
$q$=$u$ and $d$.}\label{t1}
\begin{tabular}{cccc} \hline\hline
$m_{q}$ & $m_{s}$ & $\beta_{q\bar{q}}$ & $\beta_{q\bar{s}}$ \\
\hline
0.22[0.25] & 0.45[0.48] & 0.3659[0.3194] & 0.3886[0.3419] \\
\hline\hline
\end{tabular}
\end{table}
\section{Quark Distribution Amplitudes and Decay Constants}
The quark DA of a hadron in our LFQM can be obtained from the hadronic
wave function by integrating out the transverse momenta of the quarks
in the hadron(see Eq.~(\ref{DA1})),
\bea\label{DA}
\phi(x,\mu)&=&\int^{|{\bf k}_\perp|<\mu}
\frac{d^2{\bf k}_\perp}{\sqrt{16\pi^3}}
\sqrt{\frac{\partial k_z}{\partial x}}
\Psi(x,{\bf k}_\perp,\lambda\bar{\lambda}).
\eea
For $K$ and $K^*$ meson cases, we assign the momentum fractions $x$ for 
$s$-quark and $(1-x)$ for the light $u(d)$-quark.
The quark DA describes probability amplitudes to find the hadron in a state 
with minimum number of Fock constituents and small tranverse-momentum separation
defined by an ultraviolet(UV) cutoff $\mu \gtrsim 1$GeV. 
The dependence on the scale $\mu$ is then given by the QCD
evolution equation\cite{BL} and can be calculated
perturbatively. However, the DAs at a certain low 
scale can be obtained by the necessary nonperturbative input from LFQM.
Moreover, the presence of the damping Gaussian factor in our LFQM allows
us to perform the integral up  
to infinity without loss of accuracy. The quark DAs for pseudoscalar(P)
and vector(V) mesons are constrained by 
\bea\label{DA_norm}
\int^1_0\phi_{P(V)}(x,\mu)dx=\frac{f_{P(V)}}{2\sqrt{6}},
\eea
where the decay constant is defined as 
\bea\label{fp}
\la 0|\bar{q}\gamma^\mu\gamma_5 q|P\ra=if_P P^\mu,
\eea
for a pseudoscalar meson and 
\bea\label{fv}
\la 0|\bar{q}\gamma^\mu q|V(P,\lambda)\ra&=&f_V M_V\ep^\mu(\lambda),
\nonumber\\
\la 0|\bar{q}\sigma^{\mu\nu} q|V(P,\lambda)\ra
&=&if^T_V [\ep^\mu(\lambda) P_\nu -\ep^\nu(\lambda) P_\mu],
\eea
for a vector meson with longitudinal($\lambda=0$) and 
transverse($\lambda=\pm 1$) polarizations, respectively. 
The constraint in Eq.~(\ref{DA_norm}) must be 
independent of cut-off $\mu$ up to corrections of order $\Lambda^2/\mu^2$,
where $\Lambda$ is some typical hadronic scale($\lesssim$ 1 GeV)~\cite{BL}.
For the non-perturbative valence wave
function given by Eq.~(\ref{phi}), we take 
$\mu\sim 1$ GeV as an optimal scale for our LFQM. 

The explicit form of a pseudoscalar decay constant is given by
\bea\label{fp_LFQM}
f_P=\int^1_0 dx\int [d^2{\bf k}_\perp] 
\frac{{\cal A}}{\sqrt{{\cal A}^2 + {\bf k}^2_\perp}}\phi_R(x,{\bf k}_\perp),
\eea
where ${\cal A}=(1-x)m_1 + xm_2$. The decay constants, $f_V$ and $f^T_V$,
for longitudinally and transversely polarized vector mesons, respectively, 
are given by
\begin{equation}\label{fv_LFQM}
f_V=\int^1_0 dx\int [d^2{\bf k}_\perp]
\frac{\phi_R(x,{\bf k}_\perp)}{\sqrt{{\cal A}^2 + {\bf k}^2_\perp}}
\biggl[ {\cal A} + \frac{2{\bf k}^2_\perp}{M_0+m_1+m_2}\biggr],
\end{equation} 
\begin{equation}\label{fvT_LFQM}
f^T_V=\int^1_0 dx\int [d^2{\bf k}_\perp]
\frac{\phi_R(x,{\bf k}_\perp)}{\sqrt{{\cal A}^2 + {\bf k}^2_\perp}}
\biggl[ {\cal A} + \frac{{\bf k}^2_\perp}{M_0+m_1+m_2}\biggr].
\end{equation}
The pion decay constant 
$f^{\rm exp.}_\pi\simeq 131$ MeV is meaured from 
$\pi\to\mu\nu$ and the $\rho$ meson decay constant
$f^{\rm exp.}_\rho\simeq 215$ MeV is measured from $\rho\to e^+e^-$
with the longitudinal polarization. While the constant $f_V$ are known 
from experiment, the constant $f^T_V$ are not that easily accessible in 
experiment and hence can be estimated only theoretically. 

The average value of the transverse momentum is given by
\bea\label{rmsT}
\la{\bf k}^2_\perp\ra_{Q\bar{Q}} =
\int d^3k |{\bf k}^2_\perp||\phi_R(x,{\bf k}_\perp)|^2.
\eea
Numerically, we have confirmed that 
$\la{\bf k}^2_\perp\ra^{1/2}_{Q\bar{Q}}=\beta_{Q\bar{Q}}$.
This is a 
nonperturbative measure of the transverse size in the mesonic valence
state.

We may also redefine the quark DA as 
$\Phi_{P(V)}(x)=(2\sqrt{6}/f_{P(V)})\phi(x)$ so that
\bea\label{DA2}
\int^1_0\Phi_{P(V)}(x)dx=1.
\eea
The quark DA $\Phi(x)$ evolved in the leading order(LO) of $\alpha_s(\mu)$
is usually expanded in 
Gegenbauer polynomials $C^{3/2}_n$ as
\bea\label{Gegen}
\Phi(x,\mu)&=&\Phi_{\rm as}(x)\biggl[
1+\sum_{n=1}^{\infty}a_n(\mu)C^{3/2}_n(2x-1) \biggr],
\eea
where $\Phi_{\rm as}(x)=6x(1-x)$ is the asymptotic DA and the coefficients
$a_n(\mu)$ are Gegenbauer moments~\cite{BL,Mu,Ar}.
The Gegenbauer moments with $n>0$ describe 
how much the DAs deviate from the asymptotic one. The zeroth Gegenbauer 
moment is fixed by the decay constant~\cite{BL}, e.g. for the pion:
\bea\label{Ge0}
a_0&=& 6\int^1_0[dx]\phi_\pi(x_i,\mu)=\frac{3}{\sqrt{6}}f_\pi,
\eea
where $f_\pi\simeq 131$ MeV.
In addition to the Gegenbauer moments, 
we can also define
the expectation value of the longitudinal momentum, 
so-called $\xi$-moments:
\bea\label{xi_mom}
\la\xi^n\ra&=&\int^1_{-1}d\xi \xi^n\hat{\Phi}(\xi)
=\int^{1}_{0}dx \xi^n\Phi(x),
\eea
where $\Phi(x)=2\hat{\Phi}(2x-1)$ normalized by $\la\xi^0\ra=1$.
In Appendix A, the relations between $\la\xi^n\ra$ and $a_n(\mu)$
are explicitly given up to $n=6$.
 
\section{Numerical results}
In our numerical calculations, we use two sets of the model parameters
for the linear and harmonic oscillator(HO) confining potentials  given in 
Table~\ref{t1}.  

We show in  Table~\ref{t2} our predictions for the decay
constants of $(\pi,K,\rho,K^*)$ mesons and compare with other theoretical
model predictions~\cite{BZ2,Ali} as well as data~\cite{Exp}. 
As one can see from Table~\ref{t2}, our results for the decay constants
of $(\pi,K,\rho, K^*)$ obtained from both linear and HO potential models(
especially, those from HO potential) are compatable with the 
data~\cite{Exp}.  Our values for the ratio of $f^T_V$ and $f_V$, i.e.
$f^T_\rho/f_\rho=0.76[0.80]$ and
$f^T_{K^*}/f_{K^*}=0.82[0.86]$ obtained from 
the linear[HO] potential, are quite comparable with the recent 
QCD sum rule results, $f^T_\rho/f_\rho=(0.78\pm 0.08)$
and $f^T_{K^*}/f_{K^*}=(0.78\pm 0.07)$~\cite{BZ2}.
We also find that our results for the decay constants agree surprisingly 
well with an old SU(6) symmetry
relation~\cite{Leut}, $f^T_{\rho(K^*)}=(f_{\pi(K)}+f_{\rho(K^*)})/2$ via
the sum rule $\phi^T_{\rho(K^*)}=(\phi_{\pi(K)}+\phi^L_{\rho(K^*)})/2$,
where $\phi^{L(T)}(x)$ is the longitudinally(transversely) polarized vector
meson DA. 
\begin{table}[t]
\caption{Decay constants(in MeV) for the linear[HO] potential
models compared with other models and data.}\label{t2}
\begin{tabular}{ccccc} \hline\hline
$f_M$ & Linear[HO] & SR\cite{BZ2}
& Lattice\cite{Ali} & Exp.\cite{Exp}\\
\hline
$f_\pi$ & 130[131] & - & 126.6(6.4) & 130.70(10)(36)\\
$f_K$ & 161[155] &  - & 152.0(6.1) & 159.80(1.4)(44)\\
$f_\rho$ & 246[215] & 205(9) & 239.4(7.3)
& 220(2)$^{(a)}$, 209(4)$^{(b)}$\\
$f^T_\rho$ & 188[173] & 160(10)  &- &- \\
$f_{K^*}$ & 256[223] & 217(5) & 255.5(6.5) & 217(5)$^{(c)}$\\
$f^T_{K^*}$ & 210[191] & 170(10) &- &- \\
\hline\hline
\end{tabular}

\hspace{-3.5cm}\noindent$^{(a)}$ Exp. value for $\Gamma(\rho^0\to e^+e^-)$.
\\
\hspace{-3.9cm}$^{(b)}$ Exp. value for  $\Gamma(\tau\to\rho\nu_\tau)$.
\\
\hspace{-3.6cm}$^{(c)}$ Exp. value for $\Gamma(\tau\to K^*\nu_\tau)$.
\end{table}

\begin{figure}
\includegraphics[width=2.5in,height=2.5in]{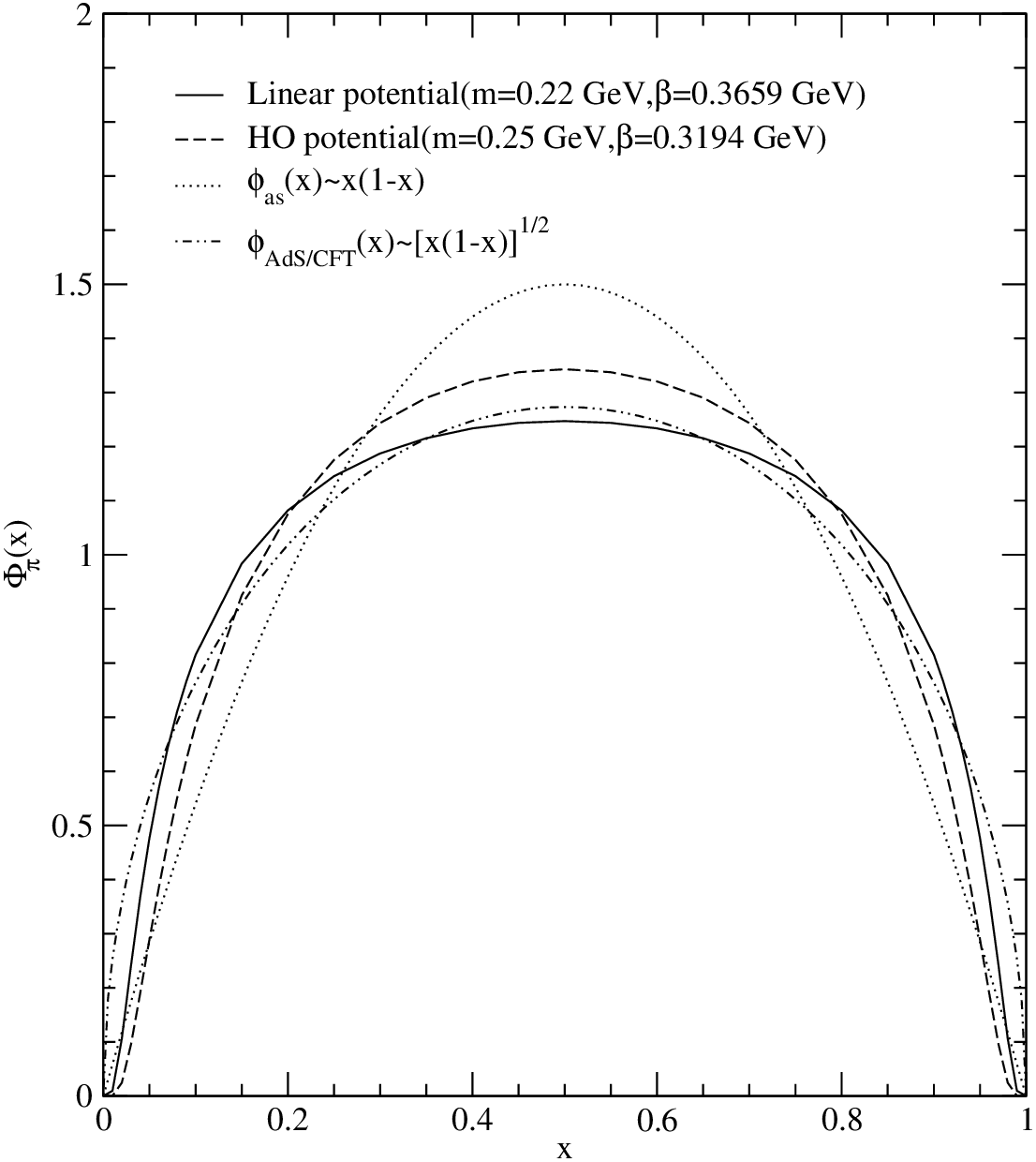}

\vspace{1.2cm}
\includegraphics[width=2.5in,height=2.5in]{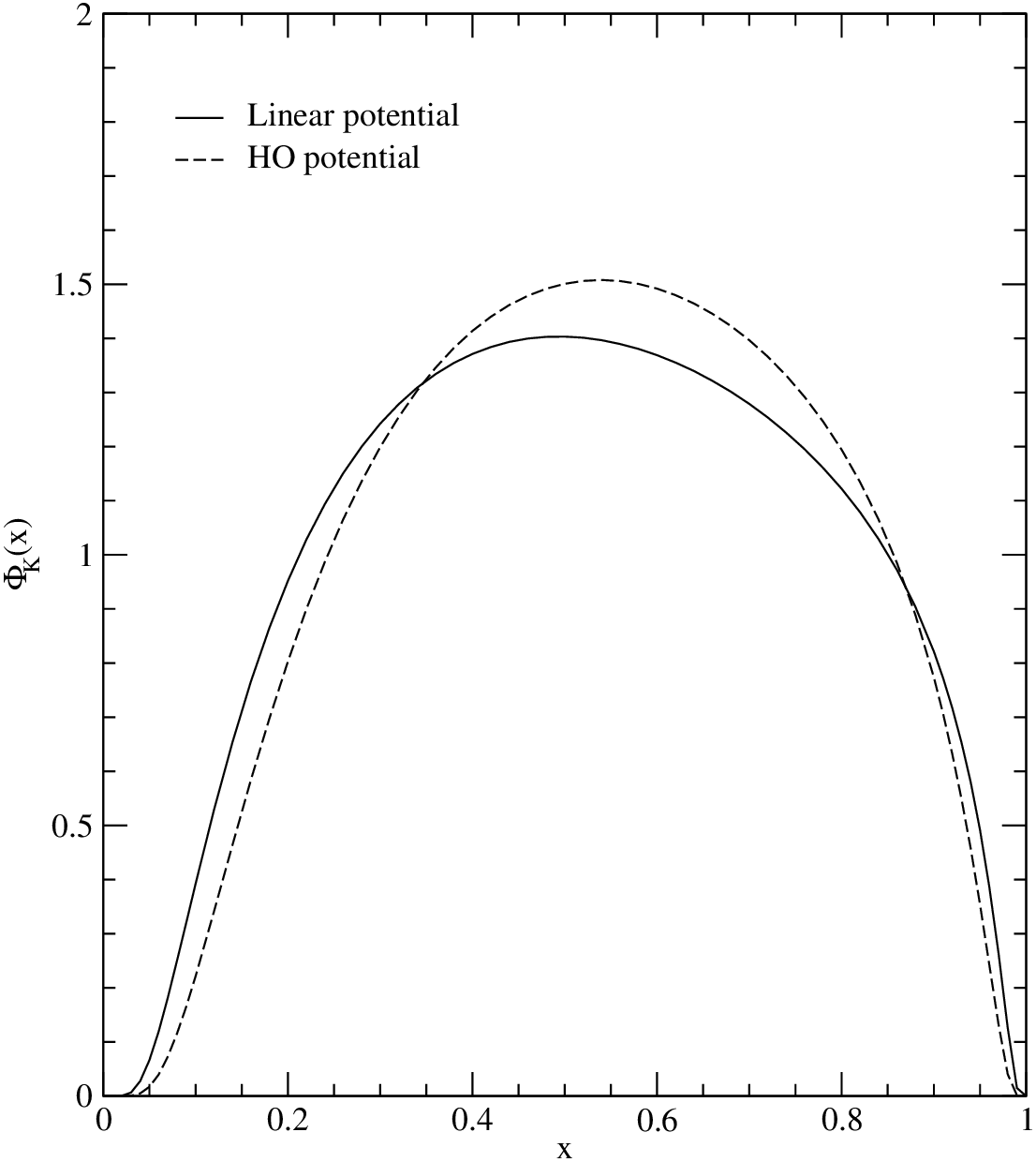}
\caption{Normalized DAs $\Phi(x)$ for $\pi$ and $K$ mesons obtained from
linear(solid line) and HO(dashed line) potential
models compared with asymptotic result(dotted line) as well as the
AdS/CFT prediction~\cite{Ads}(double-dot-dashed line).}
\label{fig1}
\end{figure}

We show in Fig.~\ref{fig1} the normalized quark DAs $\Phi_P(x)$ for $\pi$ 
and $K$ mesons obtained from linear(solid line) and HO(dashed line) 
potentials. For the pion DA, we also  compare our results with the 
asymptotic result $\Phi_{\rm as}(x)=6x(1-x)$(dotted line) as
well as the AdS/CFT prediction~\cite{Ads} 
$\Phi_{\rm AdS/CFT}(x)=\pi\sqrt{x(1-x)}/8$(double-dot-dashed line). 
For the pion case, our quark DAs
obtained from both model parameters are somewhat broader than the 
asymptotic one. We also note from the normalized pion DAs that the suppression
of the end point($x\to 0$ and 1) region has the following order in DAs,  
$\Phi_{\rm HO}(x)> \Phi_{\rm as}(x)> \Phi_{\rm Linear}(x)
> \Phi_{\rm AdS/CFT}(x)$. 
As discussed 
in Ref.~\cite{Conformal}, there exists correlation between the shape of
nonperturbative quark DA and the amount of low/high $Q^2$ contributions to
the pion form factor. As the endpoint region for the quark DA is
more suppressed, the soft(hard) contribution to the pion form 
factor gets suppressed(enhanced). 
This finding is rather similar to 
the previous findings from the Sudakov suppression of the soft
contribution\cite{LS,Jacob,JK,Doro}.
For the kaon case, the quark DA is asymmetric due to the flavor
SU(3) symmetry breaking effect. The peak points of quark DAs for two
potential models are moved slightly to the right of $x=0.5$ point
indicating that $s$-quark carries more longitudinal momentum
fraction than the light $u(d)$-quark.
\begin{figure}
\includegraphics[width=2.5in,height=2.5in]{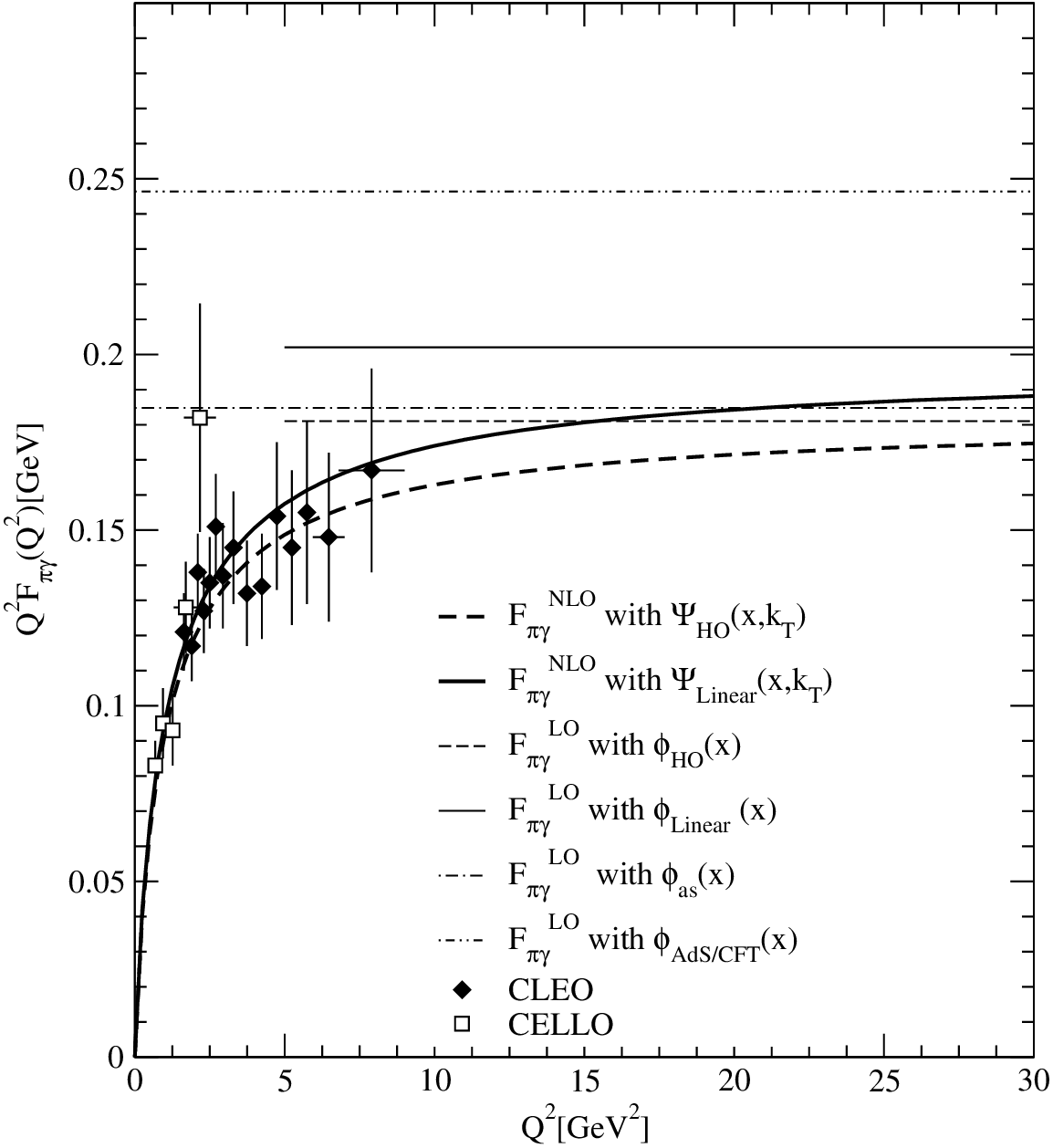}
\caption{The leading twist $F^{LO}_{\pi\gamma}(Q^2)$ and the next-to-leading
twist $F^{NLO}_{\pi\gamma}(Q^2)$ contributions to $\pi-\gamma$ transition 
form factor. Data are taken from Refs.~\cite{Gron,CELLO}.}
\label{fig2}
\end{figure}
In the LO QCD~\cite{BL}, the information of the leading-twist pion DA can be 
extracted from the pion-photon transition form factor $F_{\pi\gamma}(Q^2)$ 
as follows: 
\bea\label{trans}
\frac{Q^2F^{LO}_{\pi\gamma}(Q^2)}{\sqrt{2}f_\pi}\biggl|_{\rm twist-2}
&=&\int^1_0 dx\frac{\Phi_\pi(x,Q)}{6x(1-x)}.
\eea
The experimental value obtained in CLEO~\cite{Gron}  
is $Q^2F_{\pi\gamma}(Q^2)=(16.7\pm2.5\pm0.4)\times 10^{-2}$ GeV at
$Q^2=8$ GeV$^2$, which goes to $\sqrt{2}f_\pi\simeq 0.185$ GeV 
in the asymptotic  $Q^2\to\infty$ limit. 
With our leading twist pion DA shown in Fig.~\ref{fig1}, we obtain  
$Q^2F^{LO}_{\pi\gamma}(Q^2)= 0.202[0.181]$ GeV for the linear[HO]
potential.

For comparison between the leading twist and next-to-leading twist 
contributions to $Q^2F_{\pi\gamma}(Q^2)$, we show in Fig.~\ref{fig2}
our previous LFQM~\cite{Mixing} prediction for $Q^2F_{\pi\gamma}(Q^2)$ 
compared with the data~\cite{Gron,CELLO}.
The thick solid and thick dashed lines represent our linear and HO
potential model predictions including the higher twist effects(i.e.
${\bf k}_\perp$ and the constituent mass $m=m_u=m_d$) obtained from  
\bea\label{Fpg}
F^{NLO}_{\pi\gamma}(Q^2)&=&(e^2_u - e^2_d)\frac{\sqrt{N_c}}{\pi^{3/2}}
\int^1_0 dx\int d^2{\bf k}_\perp
\sqrt{\frac{\partial k_z}{\partial x}}
\nonumber\\
&&\times \frac{\phi_R(k^2)}{\sqrt{m^2 + {\bf k}^2_\perp}}
\frac{(1-x)m}{{\bf k'}^2_\perp + m^2},
\eea
where $N_c$ is the color factor and 
${\bf k'}_\perp = {\bf k}_\perp + (1-x){\bf q}_\perp$.
The thin solid and thin dashed lines represent our leading twist 
contribution(see Eq.~(\ref{trans})) from $\Phi_{\rm Linear}(x)$ and 
$\Phi_{\rm HO}(x)$, respectively.  We also compare our results with
the leading twist contributions from the asymptotic 
$\Phi_{\rm as}(x)$(dot-dashed line) and the AdS/CFT 
$\Phi_{\rm AdS/CFT}(x)$(double-dot-dashed line). One should note that 
the AdS/CFT prediction($\sim\sqrt{x(1-x)}$) increases the usual PQCD 
prediction($\sim x(1-x)$) by 4/3.
Our higher twist results for both potential models are not only 
very similar to each other but also in good agreement with 
the experimental data up to $Q^2\sim 10$ GeV$^2$ region.
At large $Q^2$ region, our higher twist prediction for 
the linear[HO] potential approaches 
$Q^2F^{NLO}_{\pi\gamma}(Q^2)= 0.194[0.180]$ compared to the leading twist 
result 0.202[0.181]. While the higher
twist effect on $Q^2F_{\pi\gamma}(Q^2)$ is large for the low and intermediate 
$Q^2$($\lesssim 10$ GeV$^2$) region, its effect becomes very small for 
large $Q^2$ region compared to the leading twist contribution.
Incidentally, it has been found that the leading Fock-state contribution
to $F_{\pi\gamma}(Q^2)$ fail to reproduce the $Q^2=0$ value corresponding
to the axial anomaly~\cite{anom,HW06}, i.e. it gives only a half of what
is needed to get the correct $\pi^0\to\gamma\gamma$ rate~\cite{Trei}.
However, as shown in Refs.~\cite{Ma, MR, Rad97}, the leading Fock-state
contribution to $F_{\pi\gamma}(Q^2)$ has been enhanced by replacing the 
leading Fock-state wave function to an `effective' valence quark wave function
that is normalized to one. By taking the `effective' pion wave function with
the asymptotic-like DAs, the authors in~\cite{Ma, MR, Rad97}
found an agreement with the experimental
data. Our LFQM prediction~\cite{Mixing} also uses the same approach as
Refs.~\cite{Ma, MR, Rad97}, i.e. the leading Fock-state `effective' 
wave function that is normalized to one.
The reason why our model is so 
succesful for $F_{\pi\gamma}$ transition form factor is because the $Q^2$ 
dependence($\sim 1/Q^2$) is due to the off-shell quark propagator in the 
one-loop diagram and there is no angular condition~\cite{AC} associated 
with the pseudoscalar meson.

In Tables~\ref{t3} and~\ref{t4}, we list the calculated 
$\la\xi^n\ra$ and Gegenbauer moments $a_n(\mu)$ for the pion(Table~\ref{t3}) 
and the kaon(Table~\ref{t4}) DAs obtained from the linear[HO] potential 
models at the scale $\mu\sim 1$ GeV.  We also present the comparison
with other model estimates at the scale of $1\leq\mu\leq 3$ GeV. 
While the odd Gegenbauer moments of the $\pi$ meson DA become zero due to 
isospin symmetry, the odd moments for the kaon are nonzero due to a
flavor-SU(3) violation effect of $O(m_s-m_{u(d)})$. 
For the pion case, our result for the second Gegenbauer moment,
$a^\pi_2=0.12[0.05]$ obtained from linear[HO] potential is quite comparable
with other theoretical model predictions given in Table~\ref{t3}.
A fair average is, however, $a^\pi_2=0.17\pm 0.15$ with still 
large errors~\cite{QCDSF}. The LCSR based CLEO-data
analysis~\cite{BMS1,SY,Agaev}  on the transition form factor 
$F_{\pi\gamma}$ suggests a negative value for $a^\pi_4$, 
which is consistent with the result
$a^\pi_4(1{\rm GeV}^2)>-0.07$ obtained in Ref.~\cite{BZ}. Our result 
$a^\pi_4=-0.003[-0.03]$ obtained from the linear[HO] potential also 
prefers a negative value consistent with the recent  LCSR based CLEO-data
analysis~\cite{BMS1,SY,Agaev}. 
For the kaon case, the first moment is proportional to the
difference between the longitudinal momenta of the strange and nonstrange
quark in the two-particle Fock component of the kaon, i.e.
$a^K_1=(5/3)\la x_s-x_{\bar{u}}\ra$ ($x_s=x, x_{\bar u}=1-x$).
The knowledge of $a^K_1$ is important for predicting SU(3)-violation effects
within any QCD approach that employs the quark DAs of mesons.
In our model calculation, we obtain a positive value of
$a_1^K=0.09[0.13]$ for the linear[HO] potential. 
Our results for $a^K_1$ are quite consistent with those obtained from other
estimates such as the previous LFQM~\cite{JC2}($a^K_1=0.08$), 
the chiral-quark model~\cite{NKHM}
($a^K_1=0.096$) and the QCD sum-rules~\cite{KMM,BBL}($a^K_1=0.05\pm0.02$).
The positive sign of $a^K_1$ can be understood intuitively since the heavier
strange quark(antiquark) carries a larger longitudinal momentum fraction
than the lighter nonstrange antiquark(quark).
It is interesting to note that $a^K_1$ for the HO potential model is greater 
than that for the linear one although the constituent mass difference
$m_s-m_u=0.23$ GeV is the same for both models. This difference is 
attributed to the fact that the strange quark mass for
HO potential model($m_s=0.48$ GeV) is larger than for the linear
one($m_s=0.45$ GeV) and leads to more asymmetric shape for the HO potential.  
For the second Gegenbauer moment $a^K_2$,  
however, the linear potential model gives positive value(0.03), 
while the HO one gives negative value(-0.03). 
We note that the QCD sum-rules~\cite{KMM,BBL} and 
lattice calculation~\cite{QCDSF}
give positive values while the chiral quark model~\cite{NKHM} gives 
a negative value.

\begin{figure}
\includegraphics[width=2.5in,height=2.5in]{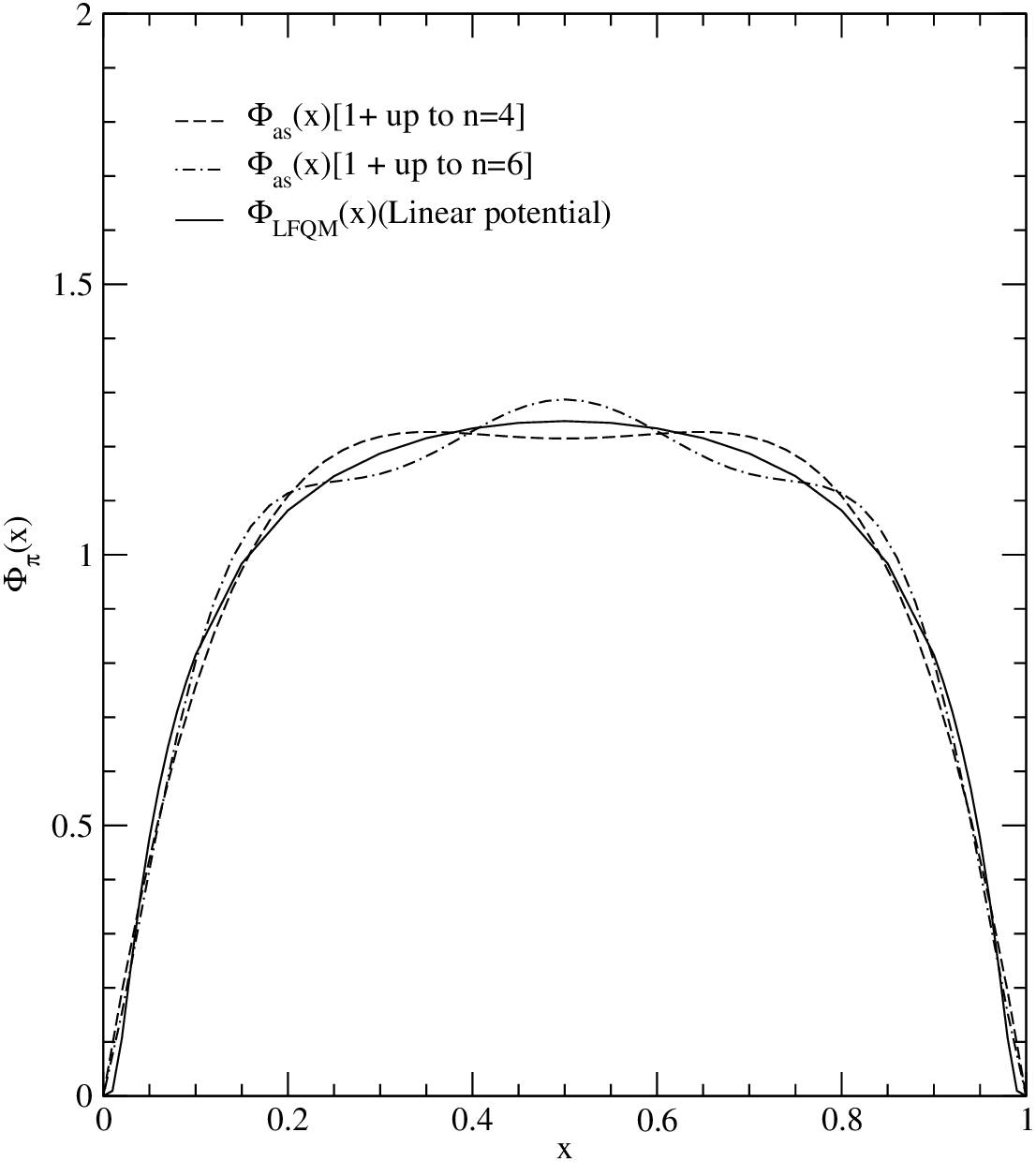}

\vspace{1.2cm}
\includegraphics[width=2.5in,height=2.5in]{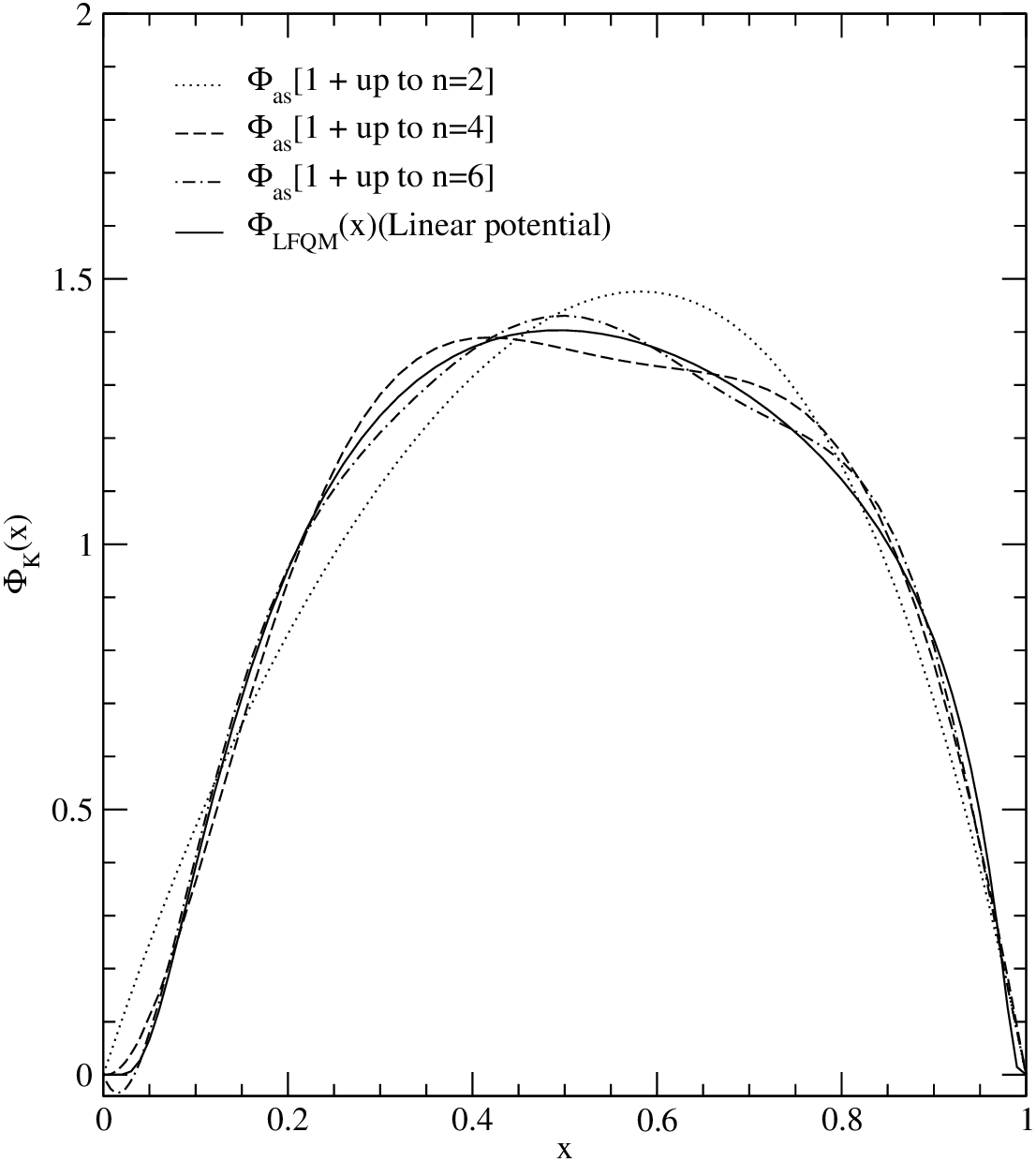}
\caption{Normalized DAs for $\pi$(upper panel) and $K$(lower panel) 
mesons obtained from linear potential model compared with those obtained 
from the truncated Gegenbauer polynomials up to $n=6$.}
\label{fig3}
\end{figure}

We show in Fig.~\ref{fig3} the normalized quark DAs for $\pi$(upper panel) 
and $K$(lower panel) mesons
obtained from the linear potential model(exact solution)
and compare with those from the truncated Gegenbauer polynomials up to
$n=6$(approximate solution). For the pion case, the truncation up to
$n=4$(dashed line) seems more close to our exact solution(solid line) than the 
truncation up to $n=6$(dot-dashed line) although the end-point behavior of 
$n=6$ case is more close to the exact solution than $n\leq 4$ case. Since 
the result up to $n=2$ truncation is not much different from that up to $n=4$ 
truncation, we do not show the result for $n=2$ case in the figure.
For the kaon case, while both truncations up to $n=4$(dashed line) and 
$n=6$(dot-dashed line) show good agreement with the exact 
solution(solid line), the truncation up to $n=2$(dotted line) 
deviates a lot from the exact solution. Thus, it seems not sufficient 
to truncate the Gegenbauer polynomials only up to $n=2$ for the kaon case.
In both cases of $\pi$ and $K$ mesons, our model calculation shows that 
the truncation of the Gegenbauer polynomials up to $n=4$ seems 
to give a reasonable approximation to the exact solution.

\begin{figure}
\includegraphics[width=2.5in,height=2.5in]{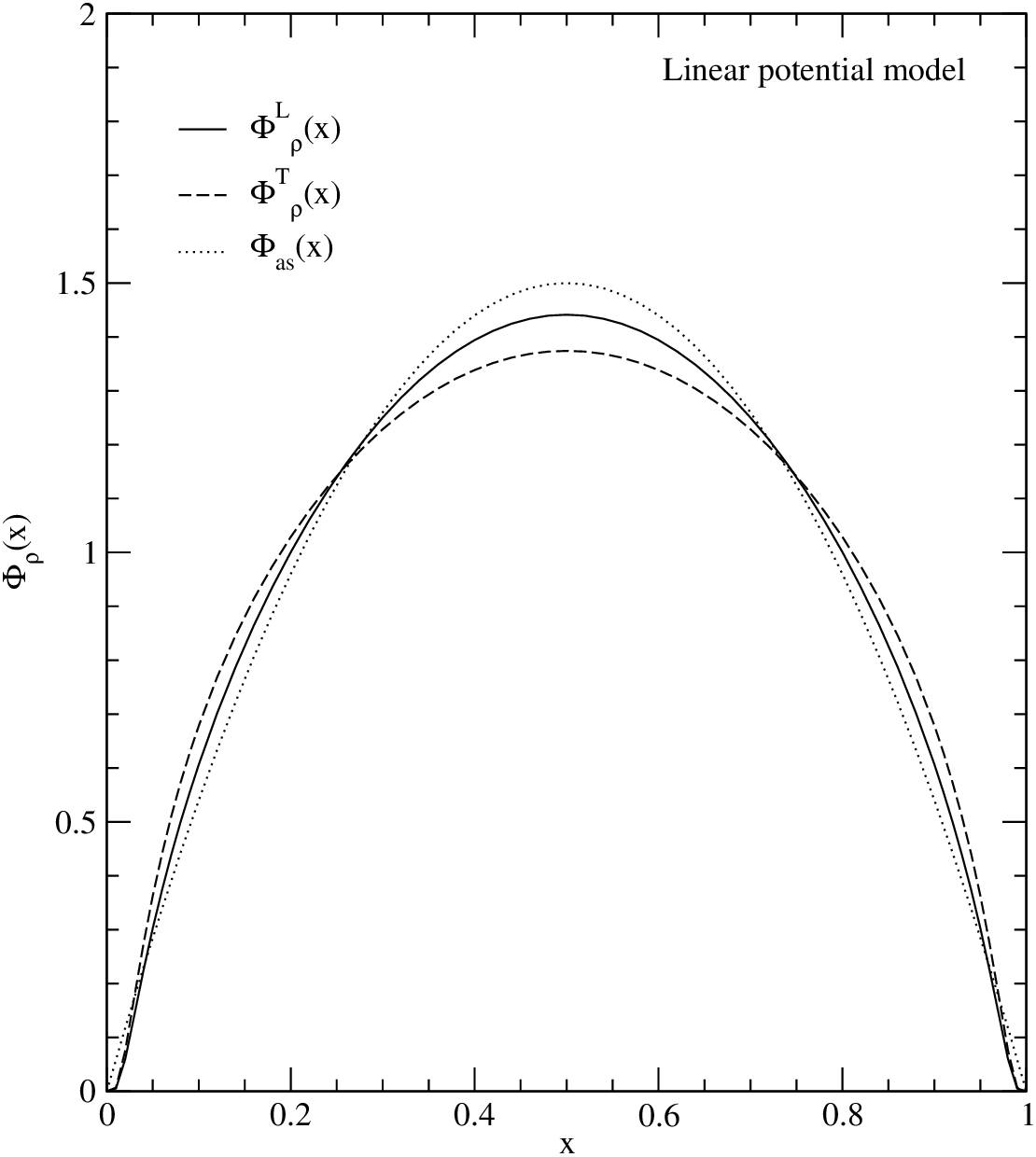}

\vspace{1.2cm}
\includegraphics[width=2.5in,height=2.5in]{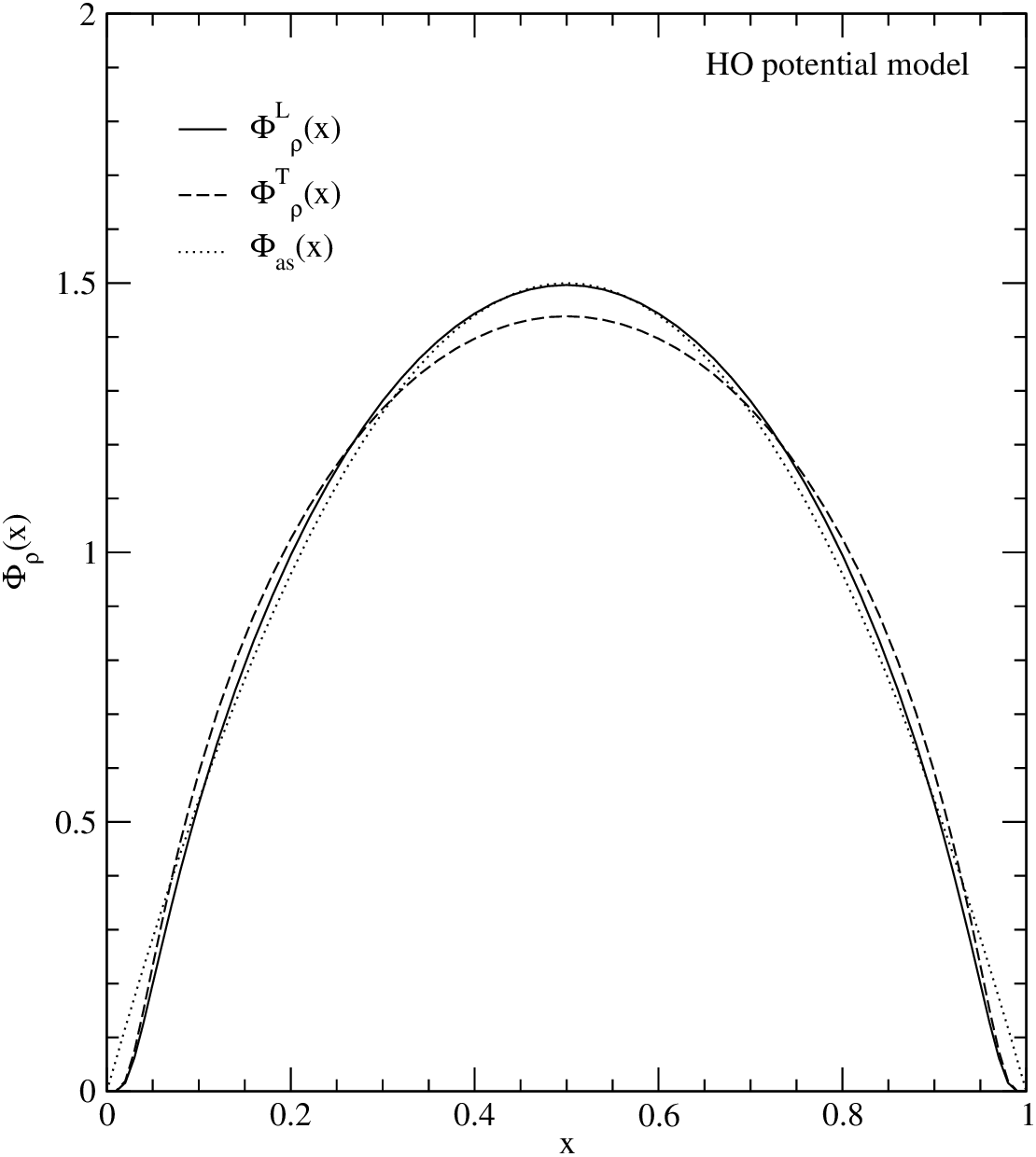}
\caption{Normalized DAs for the longitudinally(solid line) and 
transversely(dashed line) polarized $\rho$ meson obtained from
linear(upper panel) and HO(lower panel) potential models compared with
asymptotic result(dotted line).}
\label{fig4}
\end{figure}

We show in Fig.~\ref{fig4} the normalized quark DAs $\Phi_\rho(x)$
for the longitudinally(solid line) and transversely(dashed line) polarized
$\rho$ meson obtained from the linear(upper panel) and HO(lower panel)
potential models and compare with the asymptotic result(dotted line). 
For both potential models, the quark DA $\Phi^L_{\rho}$ with
longitudinal polarization is not much different from the asymptotic result and
the quark DA $\Phi^T_{\rho}$ with transverse polarization 
is somewhat broader than both $\Phi^L_{\rho}$ and $\Phi_{\rm as}$. 
Overall, the quark DAs for the $\rho$ meson are closer to the asymptotic
result than those for the pion case.
Although the overall shapes of our $\Phi^L_{\rho}$ and $\Phi^T_{\rho}$
are not much different from the asymptotic result, 
the end-point behaviors of our model 
calculation exhibiting the concave shape are different from the asymptotic 
result of the convex shape. 
We also should note that our DAs for the $\rho$ meson satisfy the
SU(6) symmetry relation~\cite{Leut},
$\Phi^T_\rho=(\Phi_\pi + \Phi^L_\rho)/2$.

In Table~\ref{t5}, we list the calculated
$\la\xi^n\ra$ and Gegenbauer moments $a_n(\mu)$ for
the $\rho$ meson DAs obtained from the linear[HO]
potential models at $\mu\sim 1$ GeV and compare with other model estimates.
The odd Gegenbauer moments of both longitudinally and transversely polarized
$\rho$ meson DAs become zero due to isospin symmetry. 
Our $\xi$ moments obtained from 
both linear and HO potential models are similar to the asymptotic
results and quite consistent with the previous LFQM~\cite{JC2}
and QCD sum-rules~\cite{BZ2}. However, slight differences of 
$\xi$ moments among different model predictions turn out to be quite 
sensitive in terms of Gegenbauer moments. For instance, the second 
Gegenbauer moment $a^\rho_{2L}$ for the longitudinally polarized $\rho$
meson obtained from the linear potential model 
gives positive value(0.02), while the same from the HO potential model 
gives negative value(-0.02). This may be still comparable with the previous 
LFQM~\cite{JC2} giving 
negative value(-0.03) and the QCD sum-rules~\cite{BZ2} giving positive 
value($0.09^{+0.10}_{-0.07}$).  Our results for the higher Gegenbauer moments
$a^\rho_n(n\geq 4)$ obtained from both linear and HO potential models 
show negative values regardless of polarization states. 
These negative values for the higher Gegenbauer moments
are related with the concave shapes of quark DAs, $\Phi^L_\rho(x)$ and
$\Phi^T_\rho(x)$, at the end-point region. 

\begin{figure}
\includegraphics[width=2.5in,height=2.5in]{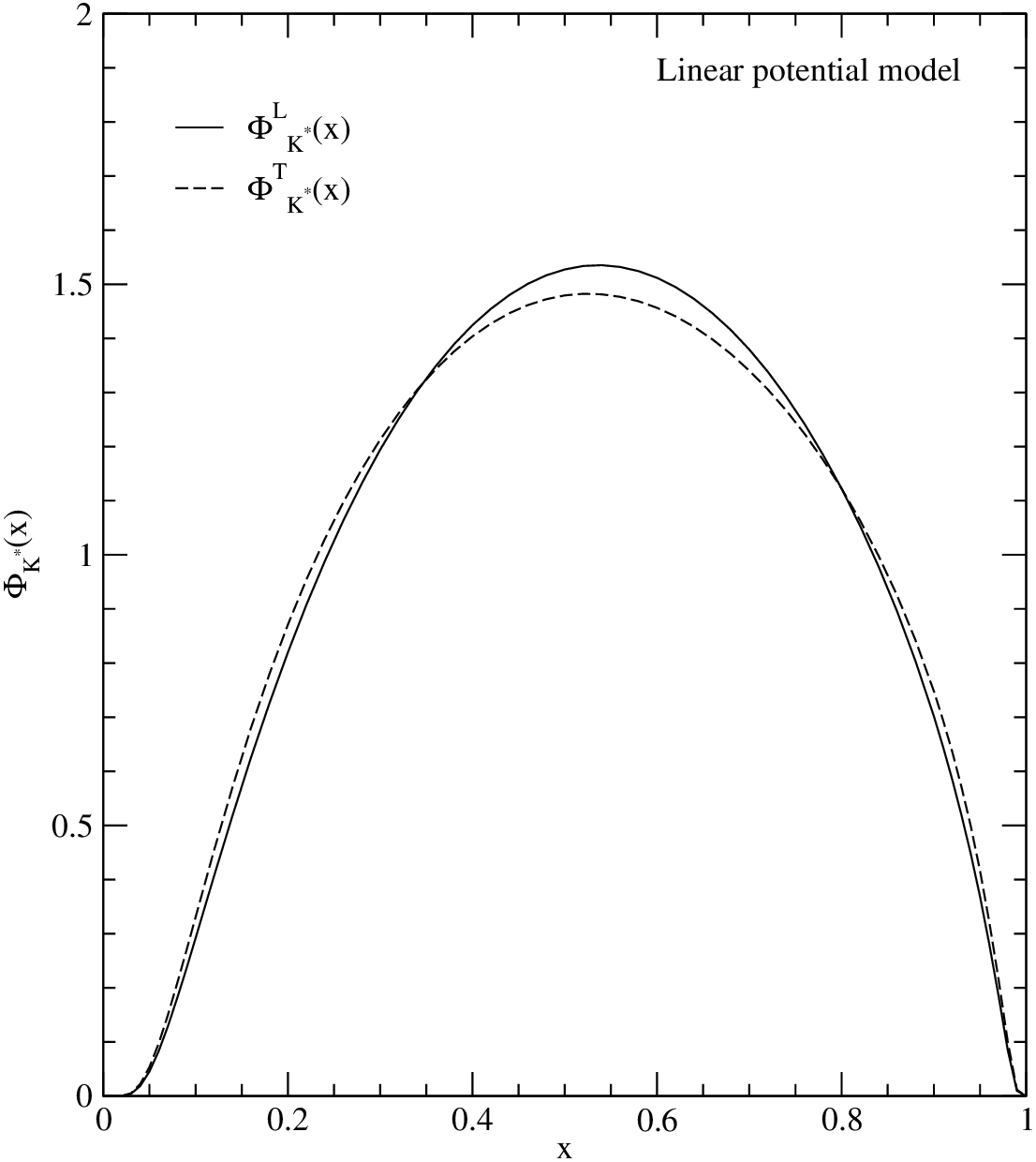}

\vspace{1.2cm}
\includegraphics[width=2.5in,height=2.5in]{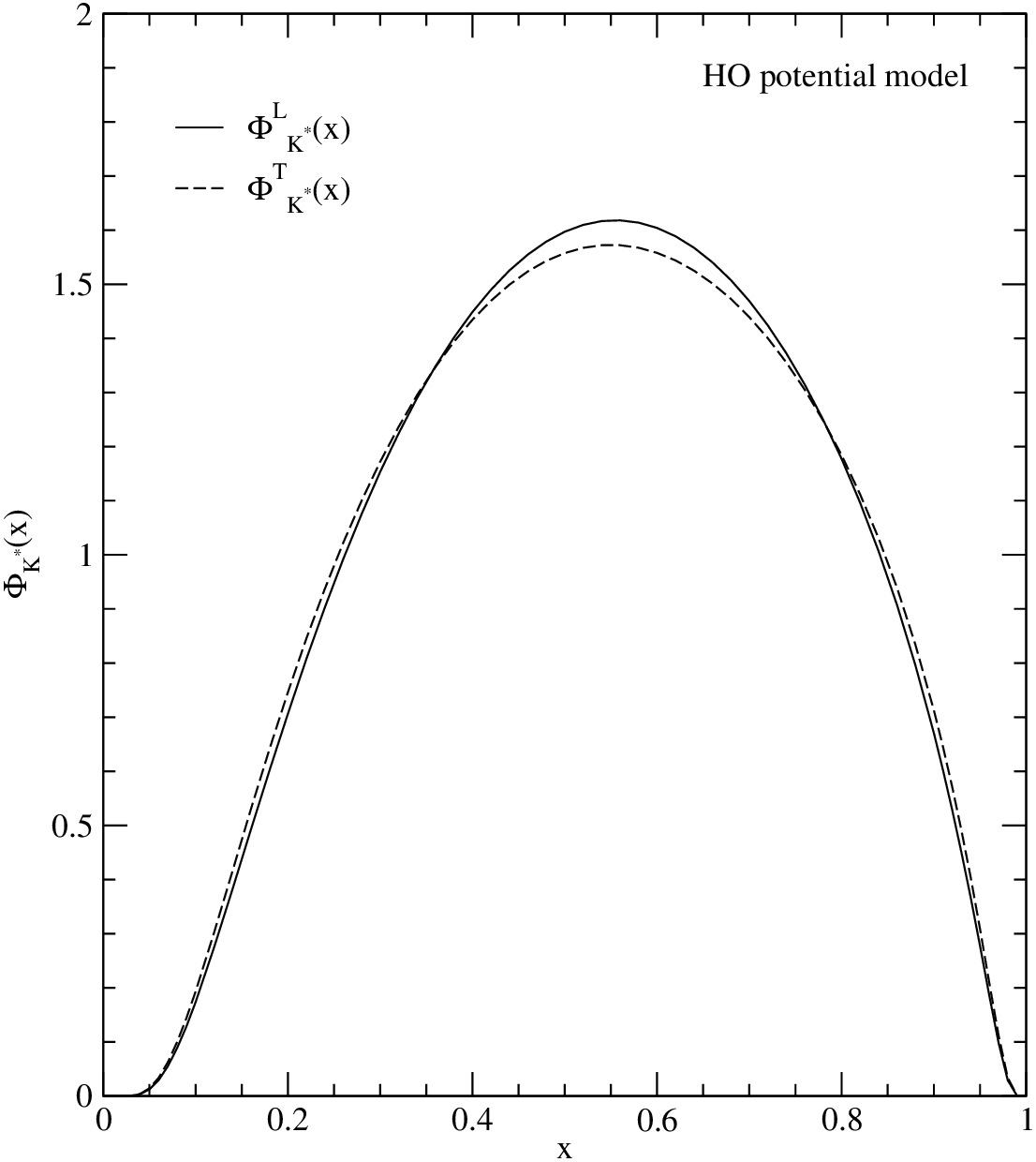}
\caption{Normalized DAs for longitudinally(solid line) and 
tranversely(dashed line) polarized $K^*$ meson obtained from
linear(upper panel) and HO(lower panel) potential models, respectively.}
\label{fig5}
\end{figure}

We show in Fig.~\ref{fig5} the normalized quark DAs $\Phi_{K^*}(x)$ for 
longitudinally(solid line) and tranversely(dashed line) polarized 
$K^*$ meson obtained from the linear(upper panel) and HO(lower panel) 
potential models. 
As in the case of the $\rho$ meson, the shape of $\Phi^T_{K^*}$ for the 
transversely polarized $K^*$ meson near the central($x=1/2$) region is 
somewhat broader than that of $\Phi^L_{K^*}$ for the longitudinally 
polarized $K^*$ meson in both linear and HO models. 
Also, the peak points for both $\Phi^L_{K^*}$ 
and $\Phi^T_{K^*}$ are shifted to the right of $x=1/2$ point due to the SU(3) 
flavor symmetry breaking as in the case of $K$ meson. 
Our quark DAs for the $K^*$ meson satisfy also the SU(6) symmetry 
relation~\cite{Leut}, $\Phi^T_{K^*}=(\Phi_K + \Phi^L_{K^*})/2$.

In Tables~\ref{t6} and~\ref{t7}, we list the calculated 
$\la\xi^n\ra$ and Gegenbauer moments $a_n(\mu)$ for 
the longitudinally(Table~\ref{t6}) and transversely(Table~\ref{t7}) 
polarized $K^*$ meson DAs obtained from the linear[HO] 
potential models 
at $\mu\sim 1$ GeV and compare with the available QCD 
sum-rule results~\cite{BZ2,Yang}.
While the odd Gegenbauer moments of the $\rho$ meson DAs become zero due to
the isospin symmetry, the odd moments for the $K^*$ meson DAs are nonzero 
because the SU(3) flavor symmetry is broken. Our values
of the first Gegenbauer moments, $a^{K^*}_{1L}=0.11[0.14]$ 
and $a^{K^*}_{1T}=0.10[0.14]$ for the linear[HO] potential model are in a
good agreement with the QCD sum-rule results~\cite{BZ2}  
$a^{K^*}_{1L}=a^{K^*}_{1T}=0.10\pm0.07$ at the scale $\mu=1$ GeV.
Note that the positive $a^{K^*}_1$ refers to $K^*$ containing an $s$ quark
but the sign will change for $\bar{K^*}$ with an $\bar{s}$ quark.
Our predictions for the even powers of $a^{K^*}_{2nL}$ and
$a^{K^*}_{2nT}(n\geq 1)$ give negative values while the LCSR results~\cite{BZ2} 
give positive values. However, the recent QCD sum-rule result 
$a^{K^*}_{3T}=0.02\pm 0.02$~\cite{Yang} at the scale $\mu=1$ GeV
is consistent with  our value $a^{K^*}_{3T}=0.04[0.03]$ obtained 
from the linear[HO] potential model. Also, the $\la\xi^{2n}\ra$ moments are not much 
different between the two models. 
\begin{figure}
\includegraphics[width=2.5in,height=2.5in]{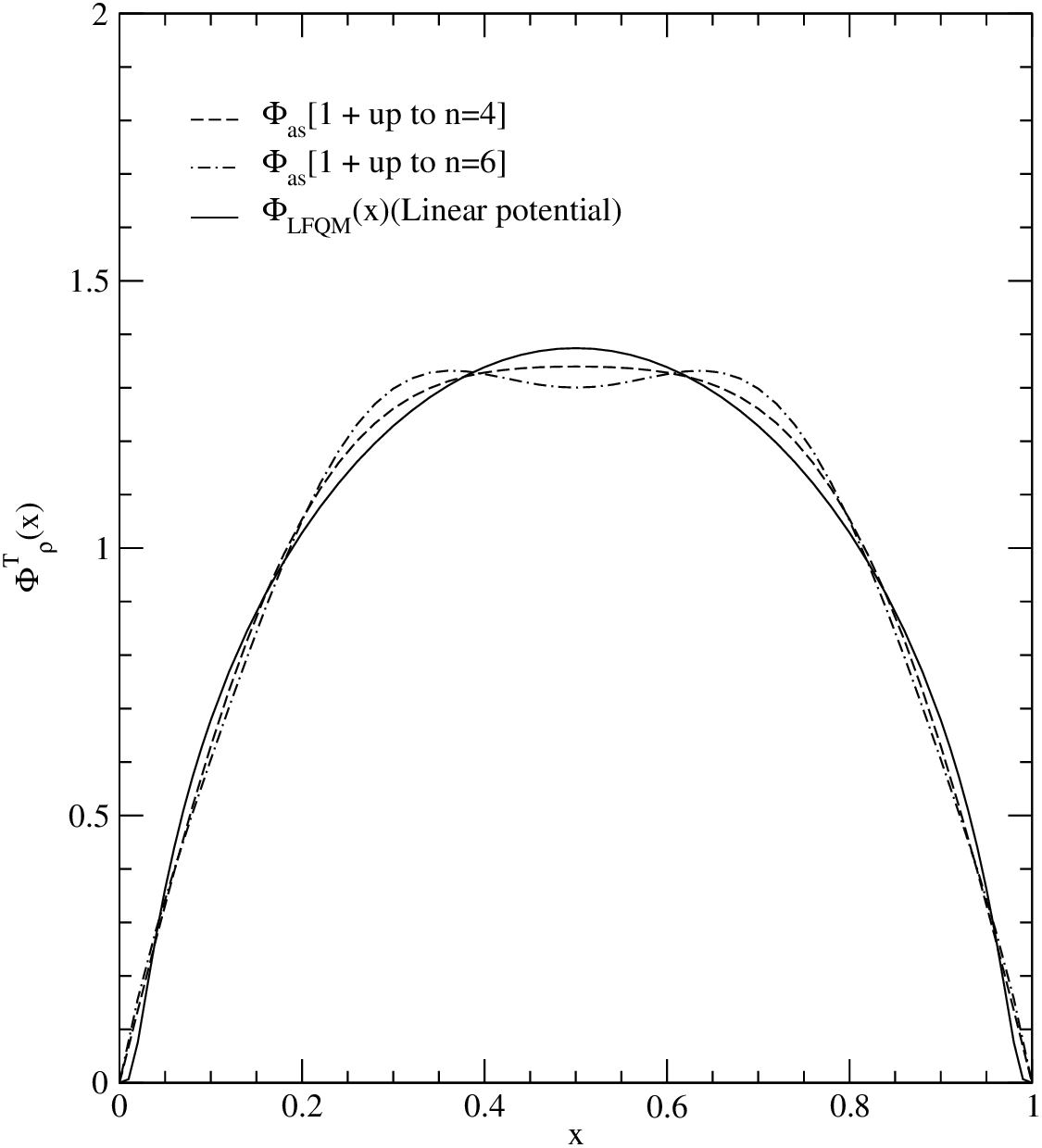}

\vspace{1.2cm}
\includegraphics[width=2.5in,height=2.5in]{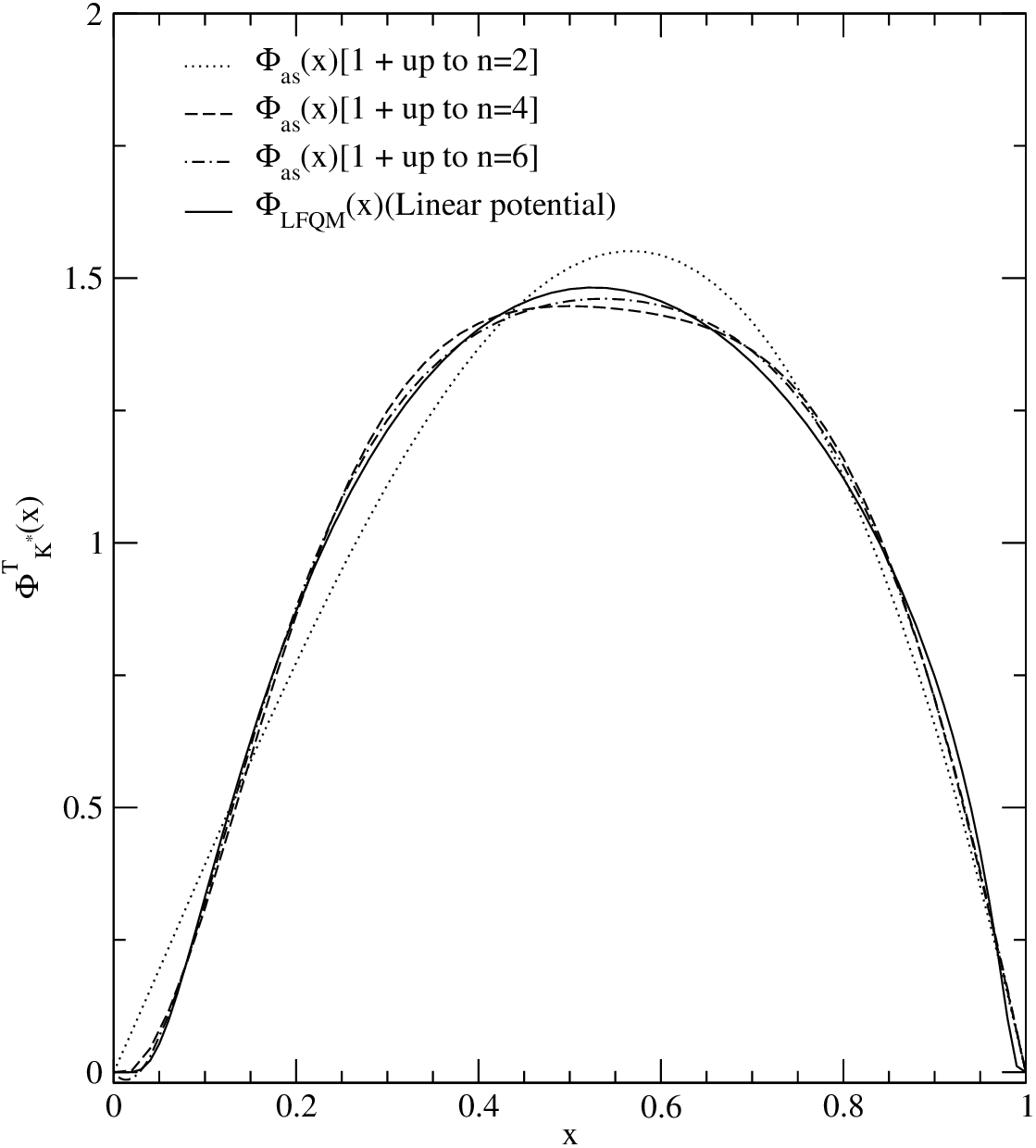}
\caption{Normalized DAs for the transversely polarized $\rho$(upper panel) 
and $K^*$(lower panel) mesons obtained from
linear potential model compared with those obtained from Gegenbauer
polynomials up to $n=6$.}
\label{fig6}
\end{figure}
We show in Fig.~\ref{fig6} the normalized quark DAs 
for transversely polarized $\rho$(upper panel) and $K^*$(lower panel) mesons
obtained from the linear potential model and compare with those from 
the truncated Gegenbauer polynomials up to $n=6$. 
For the $\rho$ meson case(upper panel), since the truncation up to
$n=2$ does not much differ from that up to $n=4$(dashed line), we do not
show the result for $n=2$ case in the figure. Both truncations up to $n=2$ 
and $n=4$ are quite close to our exact solution(solid line). 
The truncation up to $n=6$(dot-dashed line) having a deep at $x=1/2$ point 
shows a slight deviation from the exact solution. 
For the $K^*$ meson case(lower panel), both truncations up to 
$n=4$ and $n=6$ show good agreement with the exact solution, while the truncation 
up to $n=2$(dotted line) deviates a lot from the exact solution. 
For both $\rho$ and $K^*$ meson cases, 
the truncation of the Gegenbauer polynomials up to 
$n=4$ seems to give an overall reasonable approximation to the exact solution.

\section{Summary and Discussion}
In this work, we investigated the quark DAs,
the Gegenbauer moments, and decay constants for $\pi,\rho,K$ and $K^*$
mesons using the LFQM constrained by the variational principle for the 
QCD-motivated effective Hamiltonian. Our model parameters obtained from the
variational principle uniquely determine the above nonperturbative 
quantities.

Our predictions for the quark DAs for $\pi$ and $\rho$ mesons
show somewhat broader shapes than the asymptotic one. The odd Gegenbauer
moments for $\pi$ and $\rho$ meson DAs become zero due to isospin symmetry.
Our predictions for $a^\pi_2$ and $a^\pi_4$ are consistent with the recent
light-cone sum-rule based CLEO data analysis for the pion-photon transition
form factor. Interestingly, we also find that our leading twist 
result for the pion-photon transition form factor, 
$Q^2F^{LO}_{\pi\gamma}(Q^2)= 0.202[0.181]$ GeV
obtained from the linear[HO] potential model, is reduced to
$Q^2F^{NLO}_{\pi\gamma}(Q^2)= 0.194[0.180]$ if we include the
higher twist effects such as the transverse momentum and the constituent
mass. Our result is quite compatible with the CLEO data,
$Q^2F_{\pi\gamma}(Q^2)=(16.7\pm2.5\pm0.4)\times 10^{-2}$ GeV at
$Q^2=8$ GeV$^2$~\cite{Gron}.
The quark DAs for $K$ and $K^*$ show asymmetric forms due to the flavor 
SU(3)-symmetry breaking effect. This leads to the nonzero values of the
odd Gegenbauer moments.
In our model calculations of the quark DAs for ($\pi,K,\rho,K^*$) mesons,
the truncation of the Gegenbauer polynomials up to $n=4$ seems to give a
reasonable approximation to the exact solution.

Our predictions for the decay constants for $\pi$, $K$, longitudinally 
polarized $\rho$ and $K^*$ mesons are in a good agreement with the data. 
We also obtain the decay constants for the transversely polarized $\rho$ 
and $K^*$ mesons($f^T_\rho$ and $f^T_{K^*}$) using our LFQM. 
Our predicted values of $f^T_V/f_V$ averaged between the linear and HO potential
cases are $(f^T_\rho/f_\rho)_{\rm av}=0.78$ and $(f^T_{K^*}/f_{K^*})_{\rm av}=0.84$.
They are consistent with the recent QCD sum rule results,
$f^T_\rho/f_\rho=(0.78\pm 0.08)$
and $f^T_{K^*}/f_{K^*}=(0.78\pm 0.07)$~\cite{BZ2}.
Moreover, our results for the decay constants are in a good agreement with
the SU(6) symmetry relation~\cite{Leut}, 
$f^T_{\rho(K^*)}=(f_{\pi(K)}+f_{\rho(K^*)})/2$ via
the sum rule $\phi^T_{\rho(K^*)}=(\phi_{\pi(K)}+\phi_{\rho(K^*)})/2$.
Further investigations to utilize our LFQM are underway.

\acknowledgements
This work was supported by a grant from the U.S. Department of
Energy under Contract No. DE-FG02-03ER41260.  H.-M. Choi was supported in part by Korea
Research Foundation under the contract KRF-2005-070-C00039. 
This research also used resources of the National Energy Research Scientific Computing
Center, which is supported by the Office of Science of the U.S. Department of Energy
under Contract No. DE-AC02-05CH11231.

\appendix
\section{ Relation between $\xi$ and Gegenbauer moments}
\label{AppA}
The $\xi$-moments $\la\xi^n\ra$ defined by Eq.(\ref{xi_mom})
can be related to the Gegenbauer moments $a_n(\mu)$ in Eq.(\ref{Gegen}).
The relations up to $n=6$ are given by
\bea\label{ap1}
\la\xi^1\ra&=&\frac{3}{5}a_1,
\nonumber\\
\la\xi^2\ra&=&\frac{12}{35}a_2 + \frac{1}{5},
\nonumber\\
\la\xi^3\ra&=&\frac{9}{35}a_1 + \frac{4}{21}a_3,
\nonumber\\
\la\xi^4\ra&=&\frac{3}{35} + \frac{8}{35}a_2 + \frac{8}{77}a_4,
\nonumber\\
\la\xi^5\ra&=&\frac{1}{7}a_1 + \frac{40}{231}a_3 + \frac{8}{143}a_5,
\nonumber\\
\la\xi^6\ra&=&\frac{1}{21} + \frac{12}{77}a_2 + \frac{120}{1001}a_4
+ \frac{64}{2145}a_6.
\eea
Also, the first six Gegenbauer polynomials in Eq.(\ref{Gegen}) are as follows:
\bea\label{6Ge}
C^{3/2}_1(\xi) &=& 3\xi,
\nonumber\\ 
C^{3/2}_2(\xi) &=& \frac{3}{2}(5\xi^2-1),
\nonumber\\
C^{3/2}_3(\xi) &=& \frac{5}{2}\xi(7\xi^2 -3),
\nonumber\\
C^{3/2}_4(\xi) &=& \frac{15}{8}(21\xi^4 - 14\xi^2 + 1),
\nonumber\\
C^{3/2}_5(\xi) &=& \frac{21}{8}\xi(33\xi^4 - 30\xi^2 + 5),
\nonumber\\
C^{3/2}_6(\xi) &=& \frac{1}{16}(3003\xi^6 - 3465\xi^4 + 945\xi^2-35).
\eea

\newpage
\begin{table*}[t]
\caption{The $\xi$ and Gegenbauer moments $a^\pi_n(\mu)$ for the
pion DAs obtained from the linear[HO] potential
models compared with other model estimates.
The numbers in the parentheses stand for the scales of the corresponding
works.}\label{t3}
\begin{tabular}{ccccccc} \hline\hline
Models & $\la\xi^2\ra$ & $\la\xi^4\ra$ & $\la\xi^6\ra$
& $a^\pi_2$ & $a^\pi_4$ & $a^\pi_6$\\
\hline
Linear[HO] & 0.24[0.22] & 0.11[0.09] & 0.07[0.05]
& 0.12[0.05] & -0.003[-0.03] & -0.02[-0.03]\\
Asymptotic WF & 0.20 & 0.09 & 0.05 & 0 & 0 & 0\\
AdS/CFT~\cite{Ads} & 0.25 & 0.125 & 0.078 & 0.146 & 0.057 & 0.031\\
\cite{CZ}(2.4 GeV) & 0.35 & 0.21 & - & 0.44 & 0.25 & - \\
\cite{BZ}(1.0 GeV) & 0.24 & 0.11 & - & 0.115 & -0.015 & - \\
\cite{Agaev}(1.0 GeV) & 0.28 & 0.13 & - &0.23 & -0.05 & - \\
\cite{BMS2}(1.0 GeV) & 0.27 & 0.12 & - &0.20 & -0.14 & - \\
\cite{PPRWG}(1.0 GeV) & 0.22 & 0.10 & - & 0.046 & 0.007 & - \\
\cite{NKHM}(1.0 GeV) & 0.21 & 0.09 & 0.05 & 0.029 & -0.046 & -0.019\\
\cite{ADT}(1.0 GeV) & 0.22 & 0.09 & - &0.05 & -0.04 & - \\
\cite{QCDSF}(2.0 GeV) & 0.269(39) & - & - & 0.201(114)&-&-\\
\cite{SY}(2.4 GeV) & 0.24$\pm$0.01 & - & - & 0.12$\pm$0.03&-&-\\
\cite{BMS2}(2.4 GeV) & 0.25& 0.11 & - & 0.14 & -0.08 & - \\
\cite{DS}(2.4 GeV) & 0.23 & 0.11 & - & 0.08 & 0.02 & - \\
\cite{Gron}(2.4 GeV) & 0.27 & 0.11 & - & 0.19 & -0.14 & - \\
\cite{Ai}(2.4 GeV) & 0.25 & 0.12 & - & 0.16 & 0.02 & - \\
\hline\hline
\end{tabular}
\end{table*}

\begin{table*}[t]
\caption{The $\xi$ and Gegenbauer moments $a^K_n(\mu)$  for the
kaon DAs obtained from the linear[HO] potential
models compared with other model estimates.
The numbers in the parentheses stand for the scales of the corresponding
works.}\label{t4}
\begin{tabular}{ccccccc} \hline\hline
Models & $\la\xi^1\ra$ & $\la\xi^2\ra$ & $\la\xi^3\ra$
& $\la\xi^4\ra$ & $\la\xi^5\ra$ & $\la\xi^6\ra$ \\
\hline
Linear[HO]   & 0.06[0.08] & 0.21[0.19] & 0.03[0.04] & 0.09[0.08]
& 0.02[0.03] & 0.05[0.04]\\
\cite{NKHM}(1.0 GeV) & 0.057 & 0.182 & 0.023 & 0.070 & 0.012 & 0.0345\\
\cite{JC2}(1.0 GeV) & 0.046 & 0.20 & 0.025 & 0.08 & 0.015 & 0.04\\
\cite{BBL}(2.0 GeV) & 0.03$\pm$0.01& 0.26$\pm$0.04& - & - &- & -\\
\hline\hline
Models & $a^K_1$ & $a^K_2$ & $a^K_3$ & $a^K_4$ & $a^K_5$ & $a^K_6$ \\
\hline
Linear[HO] & 0.09[0.13] & 0.03[-0.03] & 0.06[0.04]
& -0.02[-0.03] & 0.007[-0.01] & -0.01[-0.01]\\
\cite{NKHM}(1.0 GeV) & 0.096 & -0.051 & -0.008 & -0.040 & -0.002 & -0.0097\\
\cite{JC2}(1.0 GeV) & 0.08 & 0 & 0.03 & -0.06 & -0.14 & -0.03\\
\cite{KMM}(1.0 GeV) & 0.05$\pm$0.02& 0.27$^{+0.37}_{-0.12}$& - & - &- & -\\
\cite{QCDSF}(2.0 GeV) &0.0453(9)(29) & 0.175(18)(47) & - & - & - & -\\
\cite{BBL}(2.0 GeV) & 0.05$\pm$0.02& 0.17$\pm$0.10& - & - &- & -\\
\hline\hline
\end{tabular}
\end{table*}

\begin{table*}
\caption{The $\xi$ and Gegenbauer moments $a^\rho_n(\mu)$ for the
$\rho$ meson DAs obtained from the linear[HO] potential
models compared with other model estimates.
The numbers in the parentheses stand for the scales of the corresponding
works.}\label{t5}
\begin{tabular}{ccccccc} \hline\hline
Models & $\la\xi^2\ra_{L}$ & $\la\xi^4\ra_{L}$ & $\la\xi^6\ra_{L}$
& $\la\xi^2\ra_{T}$ & $\la\xi^4\ra_{T}$ & $\la\xi^6\ra_{T}$ \\
\hline
Linear[HO] & 0.21[0.19] & 0.09[0.08] & 0.05[0.04]
& 0.22[0.20] & 0.10[0.09] & 0.06[0.04]\\
Asymptotic WF & 0.20 & 0.09 & 0.05 & 0.20 & 0.09 & 0.05\\
\cite{JC2}(1.0 GeV) & 0.19 & 0.07 & 0.036 & 0.2 & 0.082 & 0.042\\
\cite{BZ2}(1.0 GeV)& 0.23$^{+0.03}_{-0.02}$& 0.11$^{+0.03}_{-0.02}$ & -
&0.23$^{+0.03}_{-0.02}$ & 0.11$^{+0.03}_{-0.02}$ &-\\
\hline\hline
Models & $a^\rho_{2L}$ & $a^\rho_{4L}$ & $a^\rho_{6L}$
& $a^\rho_{2T}$ & $a^\rho_{4T}$ & $a^\rho_{6T}$ \\
\hline
Linear[HO] & 0.02[-0.02] & -0.01[-0.03] & -0.02[-0.02]
& 0.06[0.007] & -0.01[-0.03] & -0.02[-0.02]\\
\cite{JC2}(1.0 GeV) & -0.03 & -0.09 & 0.7 & 0 & -0.04 & -0.04\\
\cite{BZ2}(1.0 GeV)&0.09$^{+0.10}_{-0.07}$ & 0.03$\pm$0.02 & -
& 0.09$^{+0.10}_{-0.07}$& 0.03$\pm$0.02&-\\
\hline\hline
\end{tabular}
\end{table*}

\begin{table*}
\caption{The $\xi$ and Gegenbauer moments $a^{K^*}_n(\mu)$ for
the longitudinally polarized $K^*$ meson DAs obtained 
from the linear[HO] potential models compared with other model estimates.
The numbers in the parentheses stand for the scales of the corresponding
works.}\label{t6}
\begin{tabular}{ccccccc} \hline\hline
Models & $\la\xi^1\ra_{L}$ & $\la\xi^2\ra_{L}$ & $\la\xi^3\ra_{L}$
& $\la\xi^4\ra_{L}$ & $\la\xi^5\ra_{L}$ & $\la\xi^6\ra_{L}$ \\
\hline
Linear[HO] & 0.07[0.09] & 0.19[0.18] & 0.03[0.04]
& 0.08[0.07] & 0.02[0.02] & 0.04[0.03]\\
\cite{BZ2}(1.0 GeV) & 0.06$\pm$0.04 & 0.22$^{+0.03}_{-0.02}$ &-
& 0.10$\pm$0.02 &- &- \\
\hline\hline
Models & $a^{K^*}_{1L}$ & $a^{K^*}_{2L}$ & $a^{K^*}_{3L}$
& $a^{K^*}_{4L}$ & $a^{K^*}_{5L}$ & $a^{K^*}_{6L}$ \\
\hline
Linear[HO] & 0.11[0.14] & -0.03[-0.07] & 0.03[0.02]
& -0.02[-0.03] & 0[-0.01] & -0.01[-0.01]\\
\cite{BZ2}(1.0 GeV) & 0.10$\pm$0.07 & 0.07$^{+0.09}_{-0.07}$ &-
& 0.02$\pm$0.01 &- &- \\
\hline\hline
\end{tabular}
\end{table*}

\begin{table*}
\caption{The $\xi$ and Gegenbauer moments $a^{K^*}_n(\mu)$ for
the transversely polarized $K^*$ meson DAs obtained from the linear[HO] 
potential models and compared with other model estimates.
The numbers in the parentheses stand for the scales of the corresponding
works.}\label{t7}
\begin{tabular}{ccccccc} \hline\hline
Models & $\la\xi^1\ra_{T}$ & $\la\xi^2\ra_{T}$ & $\la\xi^3\ra_{T}$
& $\la\xi^4\ra_{T}$ & $\la\xi^5\ra_{T}$ & $\la\xi^6\ra_{T}$ \\
\hline
Linear[HO] & 0.06[0.08] & 0.20[0.18] & 0.03[0.04]
& 0.08[0.07] & 0.02[0.02] & 0.04[0.04]\\
\cite{BZ2}(1.0 GeV) & 0.06$\pm$0.04 & 0.22$^{+0.03}_{-0.02}$ &-
& 0.10$\pm$0.02 &- &- \\
\cite{Yang}(1.0 GeV) & 0.06$\pm$0.04& - & 0.03$\pm$0.02 & - & -
& - \\
\hline\hline
Models & $a^{K^*}_{1T}$ & $a^{K^*}_{2T}$ & $a^{K^*}_{3T}$
& $a^{K^*}_{4T}$ & $a^{K^*}_{5T}$ & $a^{K^*}_{6T}$ \\
\hline
Linear[HO] & 0.10[0.14] & -0.008[-0.05] & 0.04[0.03]
& -0.02[-0.03] & 0.003[-0.01] & -0.01[-0.01]\\
\cite{BZ2}(1.0 GeV) & 0.10$\pm$0.07 & 0.07$^{+0.09}_{-0.07}$ &-& 0.02$\pm$0.01
&- & - \\
\cite{Yang}(1.0 GeV) & 0.10$\pm$0.07 & - & 0.02$\pm$0.02 & - & -
& - \\
\hline\hline
\end{tabular}
\end{table*}

\end{document}